\documentclass[prd,nofootinbib,showpacs,superscriptaddress,preprint]{revtex4}
\usepackage[T1]{fontenc}
\usepackage{amsmath,amssymb}
\usepackage{epsfig}
\usepackage{dcolumn}
\usepackage{graphicx}
\usepackage[usenames,dvipsnames]{color}
\usepackage{slashed}
\usepackage[colorlinks,citecolor=blue]{hyperref}
\usepackage{pdfpages}

\begin{document}
	\title{\boldmath Phenomenology of keV scale sterile neutrino dark matter with $S_{4}$ flavor symmetry}
	
	\author{Nayana Gautam}
	\email{nayana@tezu.ernet.in}
	\affiliation{Department of Physics, Tezpur University, Tezpur - 784028, India}
	
	\author{Mrinal Kumar Das}
	\email{mkdas@tezu.ernet.in}
	\affiliation{Department of Physics, Tezpur University, Tezpur - 784028, India}

\begin{abstract}
		
We study the possibility of simultaneously addressing neutrino phenomenology and the dark matter in the framework of inverse seesaw. The model is the extension of the standard model by the addition of two right handed neutrinos and three sterile fermions which leads to a light sterile state with the mass in the keV range along with three light active neutrino states. The lightest sterile neutrino can account for a feasible dark matter(DM) candidate. We present a $S_{4}$ flavor symmetric model which is further augmented by  $Z_{4}\times Z_{3}$ symmetry to constrain the Yukawa Lagrangian. The structures of the mass matrices involved in inverse seesaw within the $S_{4}$ framework naturally give rise to correct neutrino mass matrix with non-zero reactor mixing angle $ \theta_{13}$. In this framework, we conduct a detailed numerical analysis both for normal hierarchy as well as inverted hierarchy to obtain dark matter mass and DM-active mixing which are the key factors for considering sterile neutrino as a viable dark matter candidate. We constrain the parameter space of the model from the latest cosmological bounds on the mass of the dark matter and DM-active mixing.
\end{abstract}
\pacs{12.60.-i,14.60.Pq,14.60.St}
\maketitle

\section{Introduction}
The discovery of neutrino oscillations and experimental evidences on neutrino masses and mixing provide reasons to expect physics beyond standard model. Standard model, being the most successful scenario in particle physics, can explain the mass of all the particles except neutrino through the concept of interaction of the particles with Higgs boson. Neutrinos remain massless because of the absence of right handed counter part of it in standard model (SM). The conspicuous discoveries of atmospheric and solar neutrino oscillations by the Super-Kamioka Neutrino Detection Experiment(Super-Kamiokande)\cite{superkamiokande}and
the Sudbury Neutrino Observatory (SNO)\cite{sno1,sno2}, shortly thereafter confirmed by the Kamioka Liquid Scintillator Antineutrino Detector (KamLAND)\cite{kamland} experiment  provided evidence for the neutrino flavor oscillation and massive nature of neutrinos. Some pioneer work on neutrino masses and mixing can be found in \cite{king2014neutrino,ma1998pathways,mohapatra2007theory,mass-mixingneutrino,massneutrino,massofneutrino}. In contradiction to the huge achievements in determining the neutrino oscillation parameters in solar, atmospheric, reactor and accelerator neutrino experiments\cite{abe2012reactor,1,2,abe2011indication,An2012}, many important issues related to neutrino physics  are yet not solved. Among these open questions, absolute mass scale of the neutrinos, exact nature of it (Majorana or Dirac), hierarchical pattern of the mass spectrum(Normal or Inverted), leptonic CP violation are important. The present status of different neutrino parameters can be found in global fit neutrino oscillation data\cite{OSCILLATION,oscillation2}  which is summarised in table \ref{tab3} . It is observable from the data that the leptonic mixing are much larger than the mixing in the quark sector which creates another puzzle \cite{flavor}. All these problems appeal for a coherent and successful framework beyond standard model(BSM).  

Although the results of atmospheric, solar and long-baseline experiments are well-fitted within the framework of three neutrino mixing, there are some current experiments like LSND \cite{athanassopoulos1998c},MINOS \cite{1} etc which suggest the existence of another fourth massive neutrino,usually called as sterile. LSND observed an excess of $(\bar{\nu_{e}})$ in their study of decays of positively charged muons at rest in the process  $\mu^{+}\longrightarrow e^{+} +\nu_{e} + {\bar{\nu_{\mu}}}$ while in MiniBooNE with a $\nu_{\mu}(\bar{\nu_{\mu}})$ beam having neutrino energy of $600(400)$ MeV, observed an excess of  $\nu_{e}(\bar{\nu_{e}})$ with a significance of about $4.8$ standard deviations \cite{lsnd,MiniBoone} . The results informed that muon anti-neutrinos $(\bar{\nu_{\mu}})$ converts into electron anti- neutrinos $(\bar{\nu_{e}})$ over a distance that is too short for conventional neutrino oscillations to occur. They reported that this oscillation could be explained by incorporating at least one additional neutrino. Muon anti neutrinos $(\bar{\nu_{\mu}})$ could be morphed into a fourth state of neutrino which in turn converted to electron anti neutrinos$(\bar{\nu_{e}})$. Again, anomalies in SAGE \cite{SAGE} and GALLEX \cite{gallex} have provided a hint towards the existence of fourth state of neutrinos. These additional states must relate to right-handed neutrinos (RHN) for which bare mass terms are allowed by all symmetries i.e. they should not be present in $SU(2)_{L}\times U(1)_{Y}$ interactions, hence are sterile \cite{das2019active}. Sterile neutrino as its name indicates is stable and it has only gravitational interactions. It has no charged or neutral current weak interactions except mixing with active neutrinos. Hence, the subsistence of sterile neutrinos can have observable effects in astrophysical environments and in cosmology. A natural way to produce sterile neutrinos is by their admixture to the active neutrino sector \cite{admixturesterile}.The only way to reveal the existence of sterile neutrinos in terrestrial experiments is through the effects generated by their mixing with the active neutrinos\cite{eVsterile}.

The simplest way to incorporate mass to the neutrinos is the addition of right handed (RH) neutrinos to the standard model by hand so that Higgs field can have Yukawa coupling to the neutrinos also. But to explain the smallness of neutrino mass which is in sub-eV range, the Yukawa coupling should be fine tuned to a quite unnatural value around $10^{-12}$ . Several theoretical frameworks have been proposed in the last few decades to explain tiny neutrino mass. Among all scenarios, a significant cornerstone for the theoretical research on neutrino physics is the formulation of seesaw mechanism, which is classified into different categories like Type I \cite{mohapatra1981neutrino,seesaw}, Type II \cite{arhrib2010collider}, Type III \cite{ma2002heavy,foot1989see}and Inverse Seesaw(ISS) \cite{inverse,inverse2,deppisch2005enhanced}. In Type I seesaw, the SM is extended by the addition of SM singlet fermions usually defined as right-handed(RH) neutrinos, that have Yukawa-type interactions with the SM Higgs and left-handed doublets. The light neutrino mass matrix arising from this type of seesaw is of the form $M_{\nu}\approx M_{d}M_{RR}^{-1}M_{d}^{T}$, where $M_{d}$ and $M_{RR}$ are Dirac and Majorana masses respectively. Besides the SM particle content,the type II seesaw introduces an additional $SU(2)_{L}$ triplet scalar field. Inverse seesaw requires the existence of extra singlet fermion to provide rich neutrino phenomenology. In this formalism,the lightest neutrino mass matrix is given by $M_{\nu}\approx M_{d}(M^{T})^{-1}\mu M^{-1} M_{d}^{T}$,where $M_{d}$ is the Dirac mass term and $\mu$ is the Majorana mass term for sterile, while $M$ represents the lepton number conserving interaction between right handed and sterile fermions. Neutrinos with sub-eV scale are obtained from $M_{d}$ at electroweak scale,$M$ at TeV scale and $\mu$ at keV scale. The beyond standard model(BSM) physics is successful in explaining many unperceived problems in SM as well as some other cosmological problems like baryon asymmetry of the universe(BAU) and particle nature of dark matter (DM) etc. The baryon asymmetry can be generated naturally through the decay of the RH neutrinos that exists in seesaw mechanism which is in consistent with the cosmological observable constrained by Big bang Nucleosynthesis.

Again, another important open question in particle physics as well as cosmology is dark matter(DM) and its properties. Cosmological and astrophysical measurements in different context assure that the present Universe is composed of a mysterious,non-baryonic,non-luminous matter,called dark matter \cite{DM,battaglieri2017us}. The evidence comes from the shape of cosmic microwave background(CMB)power spectrum to cluster and galactic rotation curves and gravitational lensing etc \cite{rubin1970rotation,evidence,clowe2006direct}. Inspite of these strong evidences for the presence of dark matter(DM), the fundamental nature of dark matter i.e its origin, its constituents and interactions are still unknown. According to the Planck Data , 26.8 \% of the energy density in the universe is composed of DM and the present dark matter abundance is reported as \cite{ade2016ade}:
\begin{equation*}
\Omega_{DM}h^2 = 0.1187 \pm 0.0017 
\end{equation*}

Searching for the possible connection of this exciting cosmological problem with new physics beyond standard model has been a great challenge to the physics community worldwide. The important criteria to be fulfilled by a particle to be a good DM candidate can be found in \cite{taoso2008dark}. These requirements exclude all the SM particles from being DM candidate. This has motivated the particle physics community to study different possible BSM frameworks which can give rise to the correct DM phenomenology and can also be tested at several different experiments \cite{mukherjee2017common}. There are multifarious particles that have been proposed as a remedy to the DM problem. A popular warm dark matter(WDM) candidate is sterile neutrino, a right handed neutrino, singlet under the SM gauge symmetry having tiny mixing with the SM neutrinos leading to a long lifetime \cite{abada2014looking,borah2016common}. Even in the X-ray band, one of the most studied candidates of dark matter is sterile neutrino \cite{kusenko2009sterile,abazajian2017sterile}, which can radiatively decay into an active neutrino and a mono energetic photon in the process $N\longrightarrow\nu+\gamma$ \cite{pal1982radiative}. The production of sterile neutrino DM can be naturally achieved in the early universe via a small mixing with active neutrinos \cite{dodelson1994sterile}. Warm dark matters are composed of particles having velocity dispersion between that of hot DM(HDM) and cold DM(CDM) particles. The larger free streaming length of WDM, with respect to CDM, reduces the power at small scales, suppressing the formation of small structures \cite{bode2001halo,sommer2001formation}. In general, there are no theoretical limits on the number of sterile neutrinos that can exist, on their masses and mixing with the active neutrinos \cite{eVsterile}. Sterile neutrinos specially in keV scale can have a prominent role in cosmology. Sterile neutrino is a neutral, massive particle and its lifetime can be very long as explained in \cite{merle2017kev} . The fact that sterile neutrinos can be produced by their admixture to the active neutrino sector leads to the conclusion that any reaction producing active neutrinos can also produce sterile neutrinos as long as active-sterile mixing angles are non-zero. This could produce enough sterile neutrinos to explain the observed amount of DM \cite{adhikari2017white}. However, to constitute the observed amount of DM, sterile neutrino in keV range has some constraints which can be found in \cite{adhikari2017white}. Cosmology predicts the range of viable dark matter mass as $0.4 keV \lesssim m_{DM} \lesssim 50 keV$. The lower limit is set by Tremaine-Gunn bound \cite{tremaine1979dynamical} while a DM candidate loses its stability above $50$ keV \cite{upperlimit,upper} Considering additional observational constraints on the active sterile mixing, the Doodelson-Widrow(DW) mechanism can account for atmost 50\% of the total dark matter density today \cite{abada2014dark}.

Although the origin of neutrino masses as well as leptonic mixing are not directly related to the fundamental properties of DM, it is highly inspiring to look for a common framework that can explain both the phenomena. Motivated by this, we have done a phenomenological study on the light neutrino mass matrix in the framework of inverse seesaw (2,3)(ISS(2,3)) in which the particle content of SM is extended by two right handed(RH) neutrinos and three extra sterile fermions. The motivation for studying ISS(2,3) is that it is a possible way to accommodate DM candidate within itself and also can yield correct neutrino phenomenology. Moreover, unlike canonical seesaw models, the inverse seesaw can be a low scale framework where the singlet heavy neutrinos can be at or below the TeV scale without any fine tuning of Yukawa couplings \cite{mukherjee2017common}. The symmetry realization of the model is carried out using discrete flavor symmetry $S_{4}$. The model is further augmented by $Z_{4}$ and $Z_{3}$ group to avoid the unnecessary interactions among the particles. We have constructed the matrices involved in ISS in such a way that the $\mu-\tau$ symmetry is broken in the resulting light neutrino mass matrix to comply with the experimental evidence of non-zero reactor mixing angle \cite{an2012observation}. We have studied the DM phenomenology considering the lightest sterile neutrino involved in ISS mechanism as a potential DM candidate. We have evaluated the model parameters and then feed these parameters to the calculation of DM mass, DM-active mixing, relic density and the decay rate of the lightest sterile neutrino. Our model is in agreement with the limits obtained from cosmology as well as astrophysics.

The paper is organized as follows. In section \ref{sec:level2}, we briefly review the inverse seesaw mechanism and inverse seesaw ISS (2,3) framework. In section \ref{sec:level3}, we present the $S_{4}$  flavor symmetric model and the construction of different mass matrices in lepton sector. Section \ref{sec:level4} is the discussion of dark mater production in ISS(2,3) framework. In section \ref{sec:level5},  we briefly mention the constraints on sterile neutrino dark matter. The detailed numerical analysis and the results are discussed in section \ref{sec:level6}. Finally, we conclude in section \ref{sec:level7}.
\section{\label{sec:level2}Inverse Seesaw framework : ISS(2,3)}
As discussed in several earlier works, different seesaw mechanisms have eminent role in generating tiny neutrino masses \cite{mohapatra1981neutrino,ma2002heavy,arhrib2010collider}. The inverse seesaw(ISS) incorporates an interesting alternative to the conventional seesaw. Though it requires the extension of Standard Model with extra singlet fermion,it has the advantage of lowering the mass scale to TeV range,which may be probed at LHC and in future neutrino experiments. It aims at explaining the smallness of one or more sterile neutrinos by the very same seesaw mechanism that is behind small active neutrino masses. 
The particle content in the model extends minimally that of the Standard Model, by the sequential addition of a pair of $SU(2)\times U(1)$ singlet fermions. In addition to the right-handed neutrinos,characteristic of the standard seesaw model, the inverse seesaw scheme requires gauge singlet neutrinos $s_{i}$ \cite{abada2014dark}. The relevant Lagrangian for the ISS is given as,

\begin{equation*}
L = - \dfrac{1}{2}{n_{L}^{T}C M n_{L}} + h.c
\end{equation*}
where $C \equiv i\gamma^{2}\gamma^{0}$  is the charge conjugation matrix and 
$n_{L}=(\nu_{L,\alpha},\nu_{R,i}^{c},s_{j})^{T}$.Here,$\nu_{L,\alpha}$ with $\alpha= e$,$\mu$,$\tau$ are the SM left handed neutrino states and $\nu_{R,i}^{c}(i=1,\nu_{R})$ are right handed (RH) neutrinos,while $s_{j}(j=1,s)$ are sterile fermions. The neutrino mass matrix arising from the type of Lagrangian is of the form :
\begin{equation}\label{eq:2}
M =\left(\begin{array}{ccc}
0 & M_{d}& 0\\
{M_{d}}^{T} & 0 & M_{N} \\
0 & M_{N}^{T} & \mu 
\end{array}\right) 
\end{equation}
where $M_{d}$ ,$M$ and $\mu$ are complex matrices. The matrix $\mu$ represents the Majorana mass terms for sterile fermions and the matrix M represents the lepton number conserving interactions between right handed and new sterile fermions. Standard model neutrinos at sub-eV scale are obtained from $M_{d}$ at electroweak scale, M at TeV scale and $ \mu$ at keV scale as explained in many literatures \cite{deppisch2005enhanced,dev2010tev}. The dimension of the mass matrices in the model are(\# represents the number of flavors.)
\begin{equation*}
dimension M_{d} =  (\#\nu_{L}\times \#\nu_{R})
\end{equation*}
\begin{equation*}
dimension M =  (\#\nu_{R}\times\#s)
\end{equation*}
\begin{equation*}
dimension \mu= (\#s\times \#s)
\end{equation*}
Block diagonalisation of the above mass matrix yields the effective light neutrino mass matrix as:
\begin{equation}
M_{\nu}\approx M_{d}^{T}(M_{N}^{T})^{-1}\mu M_{N}^{-1} M_{d}
\end{equation}
Again, the full $8\times8$ mass matrix can be diagolaised by using  $\mathcal{U}$ as,
\begin{equation}\label{eq:1} 
\mathcal{U}^{T}M\mathcal{U} = M^{diag} = diag(m_{1},m_{2}....m_{8})
\end{equation}
In general, the inverse seesaw formula for light neutrino mass can generate a very general structure of neutrino mass matrix. Since the leptonic mixing is found to have some specific structure with large mixing angles, one can look for possible flavor symmetry origin of it. In this context, non Abelian discrete flavor symmetries play crucial rule \cite{40discrete}.
For the purpose of the present work, we have used ISS(2,3) framework which is the extension of the SM by the addition of two RH neutrinos and three additional sterile fermions as mentioned in \cite{abada2014looking}. The motivation for using this special type of inverse seesaw is that it can account for the low energy neutrino data and also provides a viable dark matter candidate \cite{abada2014dark}. In this framework, $M_{d}$ is a $(3\times2)$ matrix,$M$ is a $(2\times3)$ matrix and $\mu$ is $(3\times3)$. The detailed mechanism for the dark matter production and the generation of light neutrino mass in the framework of ISS(2,3) will be discussed in the later sections.  The mass model is realized with the help of $S_{4}$ discrete group. The model is further augmented by $Z_{4}$ and $Z_{3}$ group to protect the unnecessary terms in the Lagrangian.
\section{\label{sec:level3}Description of the model}
Symmetries play a crucial role in particle physics. Non-Abelian discrete flavor symmetries have wide applications in particle physics as these are important tools for controlling the flavor structures of the model \cite{altarelli2010discrete,ishimori2010non}. Discrete symmetries like $A_{N}$,$S_{N}$,$Z_{N}$ are widely used in the context of model building in particle physics \cite{ma2004a_4,kalita2015constraining,zee2005obtaining,king2013neutrino,mukherjee2016neutrino,borgohain2019phenomenology}. In the present work, the symmetry realization of the structure of the mass matrices has been carried out using the discrete flavor symmetry  $S_{4}$, which is a group of permutation of four objects, isomorphic to the symmetry group of a cube.  $S_{4}$ has five irreducible representations with two singlets,one doublet and two triples denoted by $1_{1}$ , $1_{2}$, $2$, $3_{1}$ and $ 3_{2}$ respectively. The lepton doublet,charged lepton singlet of the SM and sterile fermion in inverse seesaw model transform as triplet $3_{1}$ of $S_{4}$ while the SM singlet neutrinos $N_{R}$ and SM Higgs doublet transform as $2$ and  $1_{1}$ of $S_{4}$ respectively. The product rules implemented in these representations are given in appendix \ref{appen1} . Further $Z_{4}\times Z_{3}$ symmetry is imposed to get the desired mass matrix and to constrain the non-desired interactions of the particles. The particle content and the charge assignments are detailed in table \ref{tab1}, where in addition to the lepton sector and to the SM Higgs, flavons $\phi$, $\phi^{\prime}$,$\phi_{s}$, $\chi$, $\chi^{\prime}$ have been introduced.
\begin{table}
	\centering
	\begin{tabular}{|c|c|c|c|c|c|c|c|c|c|c|c|c|c|c|}
		
		\hline 
		Field	& $\bar{L}$ & $l_{R}$ & $N_{R}$ & H & s & $\phi$ &  $\phi^{\prime}$ &$\phi_{s}$ & $\phi_{l}$ \\ 
		\hline 
		$S_{4}$ &$3_{1}$  & $3_{1}$& $2$ & $1_{1}$ & $3_{1}$& $3_{2}$ & $3_{1}$ &$1_{1}$ & $1_{1}$ \\
		\hline 
		$SU(2)_{L}$ & $2$ & $1$ &$1$& $2$ & $1$ & $1$ &  $1$ &$1$&$1$ \\
		\hline 
		$Z_{4}$& $1$ & $1$ & $1$ &$+i$& $+i$ & $-i$ & $-i$ &  $-1$ &$-i$ \\
		\hline 
		$Z_{3}$& $\omega^{2}$ & $1$ &$1$& $1$ & $\omega$ & $\omega$ &  $\omega$ &$\omega$ & $\omega$ \\
		\hline 
	\end{tabular} 
	\caption{Fields and their respective transformations
		under the symmetry group of the model.} \label{tab1}
\end{table}
The Yukawa Lagrangian for the charged leptons and also for the neutrinos can be expressed as:
\begin{equation}
-\mathcal{L}  = \mathcal{L}_{\mathcal{M_{L}}}+\mathcal{L}_{\mathcal{M_{D}}} + \mathcal{L}_{\mathcal{M}}+\mathcal{L}_{\mathcal{M_{S}}}+ h.c
\end{equation}
where,
\begin{equation}\label{eq:3}
\mathcal{L}_{\mathcal{M_{D}}} =  \frac{y}{\Lambda}\bar{L}N_{R}H\phi +  \frac{y\prime}{\Lambda}\bar{L}N_{R}H\phi^{\prime},
\end{equation}
$\Lambda$ is the cut-off scale. It is needed to lower the mass dimension to $4$. Here, we need extra scalar $\phi$ and $\phi^{\prime}$ with SM Higgs because $S_{4}$ product between a doublet and triplet yields two triplets and H is a $S_{4}$ singlet, so we require another triplet scalar so that the total Lagrangian remains singlet under $S_{4}$. 
\begin{equation}
\mathcal{L}_{\mathcal{M_{S}}} = y_{s}ss\phi_{s}
\end{equation}
$ \mathcal{L}_{\mathcal{M_{L}}}$ is the Lagrangian for the charged leptons which can be written in terms of dimension five operators as \cite{leptonmatrix}
\begin{equation}
\mathcal{L}_{\mathcal{M_{L}}} = \frac{y_{l}}{\Lambda}\bar{L}l_{R}H\phi_{l} 
\end{equation}
The following flavon alignments allow us to have the desired mass matrix corresponding to the charged lepton sector 
$$\langle \Phi_{l} \rangle = v_{l}.$$
The charged lepton mass matrix is then given by,
\begin{equation}\label{eq:g}
m^0_{l}= \frac{v_{h}}{\Lambda}\left(\begin{array}{ccc}
y_{l}v_{l} & 0& 
0 \\
0 & y_{l}v_{l} &
0\\ 
0& 0 & y_{l}v_{l}
\end{array}\right),
\end{equation}
The Lagrangian $\mathcal{L}_{\mathcal{M}}$ is given by,
\begin{equation}
\mathcal{L}_{\mathcal{M}} = y_{r} N_{R}s{\chi}+y_{r^{\prime}}N_{R}s\chi^{\prime}
\end{equation} 
The charge assignments that have been chosen for the new flavon fields $\chi$ and $\chi^{\prime}$ are shown in table \ref{tab2}.
\begin{table}
	\centering
	\begin{tabular}{|c|c|c|}
		\hline 
		Field  & $\chi$ & $\chi^{\prime}$\\ 
		\hline 
		$S_{4}$ &$3_{2}$  & $3_{1}$\\
		\hline
		$SU(2)_{L}$ & $1$ & $1$\\
		\hline 
		$Z_{4}$ & $-i $ & $-i$ \\
		\hline 
		$Z_{3}$& $\omega^{2}$ & $\omega^{2}$\\
		\hline
	\end{tabular} 
	\caption{New flavon fields with their respective transformations
		under the symmetry group of the model.}\label{tab2}
\end{table}
The VEV alignments for the flavons which result in desired neutrino mass matrix and leptonic mixing matrix are followed as:
$$\langle \Phi \rangle =(v_{h1},v_{h2},-v_{h2}), \; \langle \phi^{\prime} \rangle = (v_{h1^{\prime}},v_{h2^{\prime}},v_{h2^{\prime}}), \;\langle \Phi_{s} \rangle = v_{s}, \; \langle H \rangle =  v_{h},\; \langle \chi \rangle = (v_{r},v_{{r}\prime},0),\; \langle \chi^{\prime} \rangle = (0,v_{{r}\prime\prime},0)$$
With these flavon alignments different terms in the Lagrangian given by equation \eqref{eq:3}  can be written as;
\begin{align}
\frac{y}{\Lambda} N_{iR}L_{j}H\Phi &= \frac{y}{\Lambda}H [N_{1R}L_{1}\Phi_{1} + (\frac{\sqrt{3}}{2}N_{2R}L_{2} -\frac{1}{2} N_{1R}L_{2})\Phi_{2} + (-\frac{\sqrt{3}}{2}N_{2R}L_{3} -\frac{1}{2} N_{1R}L_{3})\Phi_{3}] \nonumber \\
& = \frac{y v_{h}}{\Lambda}[N_{1R}L_{1}v_{h1} + (\frac{\sqrt{3}}{2}N_{2R}L_{2} -\frac{1}{2} N_{1R}L_{2})v_{h2}+(\frac{\sqrt{3}}{2}N_{2R}L_{3}+\frac{1}{2} N_{1R}L_{3})v_{h2}]
\end{align}
\begin{align}
\frac{y\prime}{\Lambda} N_{iR}L_{j}H\Phi^{\prime} &=\frac{y\prime}{\Lambda}H [N_{2R}L_{1}{\Phi_{1}}^{\prime} + (-\frac{\sqrt{3}}{2}N_{1R}L_{2} -\frac{1}{2} N_{2R}L_{2}){\Phi_{2}}^{\prime} + (\frac{\sqrt{3}}{2}N_{1R}L_{3} -\frac{1}{2} N_{2R}L_{3}){\Phi_{3}}^{\prime}]\nonumber \\
& =\frac{y^{\prime}v_{h}}{\Lambda} [N_{2R}L_{1}v_{h1^{\prime}} + (-\frac{\sqrt{3}}{2}N_{1R}L_{2}-\frac{1}{2} N_{2R}L_{2})v_{h2^{\prime}}+ (\frac{\sqrt{3}}{2}N_{1R}L_{3}-\frac{1}{2} N_{2R}L_{3})v_{h2^{\prime}}]
\end{align}  
\begin{align}
y_{r} N_{iR}s_{j}\chi &= y_{r} [N_{1R}s_{1}\Phi_{1R} + (\frac{\sqrt{3}}{2}N_{2R}s_{2} -\frac{1}{2} N_{1R}s_{2})\Phi_{2R} + (-\frac{\sqrt{3}}{2}N_{2R}s_{3} -\frac{1}{2} N_{1R}s_{3})\Phi_{3R}] \nonumber \\
& = y_{r}[(N_{1R}s_{1})v_{r} + (\frac{\sqrt{3}}{2}N_{2R}s_{2} -\frac{1}{2} N_{1R}s_{2})v_{r^{\prime}}]
\end{align}
\begin{align}
{y_{r}}^{\prime} N_{iR}s_{j}\chi^{\prime} &={y^{\prime}}_{r} [N_{2R}s_{1}{\Phi_{1R}}^{\prime} + (-\frac{\sqrt{3}}{2}N_{1R}s_{2} -\frac{1}{2} N_{2R}s_{2}){\Phi_{2R}}^{\prime} + (\frac{\sqrt{3}}{2}N_{1R}s_{3} -\frac{1}{2} N_{2R}s_{3}){\Phi_{3R}}^{\prime}]\nonumber \\
& = {y_{r}}^{\prime}[(-\frac{\sqrt{3}}{2}N_{1R}s_{2} -\frac{1}{2} N_{2R}s_{2})v_{r^{\prime\prime}}]
\end{align}
\begin{equation}
y_{s} ss \Phi_{s} = y_{s} (s_{1}s_{1} + s_{2}s_{2} + s_{3}s_{3})v_{s}.
\end{equation} 
The chosen flavon alignments lead to different matrices involved in inverse seesaw formula as follows,
\[
M_{d} =\frac{v_{h}}{\Lambda}\begin{pmatrix}
y v_{h1} & y^{\prime} v_{h1^{\prime}} \\
-\frac{1}{2}y v_{h2}-\frac{\sqrt{3}}{2}y^{\prime} v_{h2^{\prime}} & \frac{\sqrt{3}}{2}y v_{h2}-\frac{1}{2}y^{\prime} v_{h2^{\prime}}\\
\frac{1}{2}y v_{h2}+\frac{\sqrt{3}}{2}y^{\prime} v_{h2^{\prime}} &  \frac{\sqrt{3}}{2}y v_{h2}-\frac{1}{2}y^{\prime} v_{h2^{\prime}}\\
\end{pmatrix}
\]
\[
M = \begin{pmatrix}
y_{r} v_{r} & -\frac{\sqrt{3}}{2}y_{r^{\prime}} v_{r^{\prime\prime}}-\frac{1}{2}y_{r}v_{r^{\prime}} & 0 \\
0 & \frac{\sqrt{3}}{2}y_{r} v_{r\prime}-\frac{1}{2}y_{r^{\prime}} v_{r^{\prime\prime}} & 0\\
\end{pmatrix}
\]
\[\mu  = y_{s}\begin{pmatrix}
1 & 0 & 0 \\
0 & 1 & 0\\ 
0 & 0 & 1
\end{pmatrix}v_{s}\]
Now,we denote $a = \frac{y}{\Lambda}v_{h} v_{h1}$ ,  $b = \frac{v_{h}}{\Lambda} (-\frac{1}{2}y v_{h2}-\frac{\sqrt{3}}{2}y^{\prime} v_{h2^{\prime}})$,  $c = \frac{v_{h}}{\Lambda}(\frac{\sqrt{3}}{2}y v_{h2}-\frac{1}{2}y^{\prime} v_{h2^{\prime}})$,  
$e = \frac{y^{\prime}}{\Lambda}v_{h} v_{h1^{\prime}}$, $f = y_{r} v_{r}$ ,$h =-\frac{\sqrt{3}}{2}y_{r^{\prime}} v_{r^{\prime\prime}}-\frac{1}{2}y_{r}v_{r\prime}$, $g = \frac{\sqrt{3}}{2}y_{r} v_{r\prime}-\frac{1}{2}y_{r^{\prime}} v_{r^{\prime\prime}}$, $p =y_{s} v_{s}$. With these notations the mass matrices involved in ISS can be written as,
\begin{equation} \label{eq:u} 
M_{D}= \left(\begin{array}{ccc}
a & e \\
b & c \\ 
-b & c 
\end{array}\right),\;  \mu  = \left(\begin{array}{ccc}
p & 0 & 0 \\
0 & p & 0\\ 
0 & 0 & p
\end{array}\right), \;  M =\left(\begin{array}{ccc}
f & h & 0 \\
0 & g & 0 
\end{array}\right).
\end{equation}
The elements of the light neutrino mass matrix in the framework of ISS(2,3)  arising from the above mentioned mass matrices are given below:
\begin{equation*}
(-m_{\nu})_{11} = \frac{(a^{2}g^{2}-2aegh+e^{2}(f^{2}+h^{2}))p}{f^{2}g^{2}}
\end{equation*}
\begin{equation*}
(-m_{\nu})_{12} =\frac{(-acgh+bg(ag-eh)+ce(f^{2}+h^{2}))p}{f^{2}g^{2}}
\end{equation*}
\begin{equation*}
(-m_{\nu})_{13} = \frac{(-acgh+bg(-ag+eh)+ce(f^{2}+h^{2}))p}{f^{2}g^{2}}
\end{equation*}
\begin{equation*}
(-m_{\nu})_{22} =\frac{(b^{2}g^{2}-2bcgh+c^{2}(f^{2}+h^{2}))p}{f^{2}g^{2}}
\end{equation*}
\begin{equation*}
(-m_{\nu})_{23} = \frac{(b^{2}g^{2}-c^{2}(f^{2}+h^{2}))p}{f^{2}g^{2}}
\end{equation*}
\begin{equation*}
(-m_{\nu})_{33} = \frac{(b^{2}g^{2}+2bcgh+c^{2}(f^{2}+h^{2}))p}{f^{2}g^{2}}
\end{equation*}

Here, we have used $S_{4}$ product rules in formulating the mass matrices that are discussed above. The light neutrino mass matrix obtained here in both the models can give rise to the correct mass squared difference and non-zero $\theta_{13}$. Besides, to avoid some non-desired interactions among the fields we require  $Z_{4}\times Z_{3}$ symmetry. Thus the desired structures of the mass matrices have been made possible by the combination of flavor symmetry $S_{4}$ as well as  $Z_{4}\times Z_{3}$ symmetry.
\section{\label{sec:level4}Dark Matter in the ISS(2,3) framework}
As mentioned in the previous sections, ISS(2,3) is a common framework to address neutrino phenomenology as well as the cosmological DM problem. This type of inverse seesaw bears the specialty that there exists an additional intermediate state with mass $m_{s}= \mathcal{O}(\mu)$ in the mass spectrum.  

Depending on the number of fields in the model, a generic ISS realization is characterized by the following mass spectrum \cite{abada2014looking}

1. Three light active states with masses of the form 
\begin{equation*}
m_{\nu} = \mathcal{O}(\mu)\frac{k^{2}}{1+k^{2}}  , k = \frac{\mathcal{O}(M_{d})}{\mathcal{O}(M_{N})}
\end{equation*}

2. \#$\nu_{R}$ pairs of pseudo-Dirac heavy neutrinos with masses$\mathcal{O}(M)$ + $\mathcal{O}(M_{d})$.

3. \#s--\#$\nu_{R}$ light sterile states (present only if \#s \textgreater\#$\nu_{R}$) with masses $\mathcal{O}(\mu)$.

In order to realize the model phenomenologically, after diagonalizing the matrix associated to a given ISS realization, must produce three light active eigenstates with mass differences in agreement with oscillation data and a mixing pattern compatible with the experimental data. In the scenario of ISS if \#s \textgreater\#$\nu_{R}$,there exists a light sterile state with mass $\mathcal{O}(\mu)$. This state has unique feature that if the mass of
this is in the range of several keV and with very small mixing with the active neutrinos, it will provide a particle with the lifetime exceeding the age of the Universe, which can be a WDM candidate \cite{merle2017kev,adhikari2017white}. ISS has also the advantage that it allows for a non zero mixing between the active and the additional sterile states. \cite{abada2017neutrino} . Among different ISS scenario, ISS(2,3) is considered to be the lowest possible way to accommodate a viable dark matter candidate \cite{abada2014looking,abada2014dark} as it contains unequal number of  $N_{R}$ and  $s$, leading to a DM particle as well as two heavy pseudo-Dirac pairs \cite{lucente2016implication}.

Now, the mass matrix in this framework with $\|\mu\|\leqslant\|M_{d}\|,\|M_{N}\|$ can be block diagonalised into light and heavy sectors as \cite{awasthi2013neutrinoless}
\begin{equation}\label{eq:a} 
m_{\nu}\approx M_{d}(M_{N}^{T})^{-1}\mu M_{N}^{-1}M_{d}^{T} 
\end{equation}
\[
M_{H} = \begin{pmatrix}
0 & M_{N} \\
M_{N}^{T} & \mu \\
\end{pmatrix}
\]
where $m_{\nu}$ is the famous ISS formula and $M_{H}$ is the mass matrix for the heavy pseudo-Dirac pairs and the extra state. $m_{\nu}$ is diagonalised by PMNS $U_{\nu}$ to get the light active neutrinos. While the diagonalisation of $M_{H}$ will give the mass of the other five heavy particles.\\
In the framework of ISS(2,3), $M$ is not a squared matrix rather it is a $2\times3$ matrix. So, $M^{-1}$ is not well defined. We followed the general version of \eqref{eq:a}as proposed by the authors \cite{abada2017neutrino} . It follows as,
\begin{equation}\label{eq:20}
m_{\nu}\approx M_{d}dM_{d}^{T} 
\end{equation} 
where d is $2\times2$ dimensional submatrix defined as
\[
{M_{H}}^{-1} = \begin{pmatrix}
d_{2\times2} & .... \\
..... &.... \\
\end{pmatrix}
\]
with 
\[
M_{H} = \begin{pmatrix}
0 & M_{N}\\
M_{N}^{T} & \mu \\
\end{pmatrix}
\]
It is important to calculate the active-sterile mixing to study the DM phenomenology. In this context, 
one can obtain the active-sterile mixing from the first three components of the eigenvectors corresponding to the keV ranged eigenvalue of the mass matrix M in equation
 \eqref{eq:2}.
Any stable neutrino state with a non-vanishing mixing to the active neutrinos will be produced through active-sterile neutrino conversion and the resulting relic abundance is proportional to the active-sterile mixing and the mass and can be expressed as \cite{asaka2007lightest}:
\begin{equation}\label{eq:c}
\Omega_{DM}h^{2} = 1.1 \times 10^{7}\sum C_{\alpha}(m_{s})|\mathcal{U}_{\alpha s}|^{2}{\left(\frac{m_{s}}{keV}\right)}^{2},  \alpha = e,\mu,\tau
\end{equation}
The derivation of the above formula can be found in appendix \ref{appen2} . It can be simplified to the following expression,
\begin{equation}
\Omega_{DM}h^{2}\approxeq0.3 {\left(\frac{sin^{2}2\theta}{10^{-10}}\right)}{\left(\frac{m_{s}}{100keV}\right)}^{2}
\end{equation}
where $sin^{2}2\theta = 4\sum_{\alpha = e,\mu,\tau} |\mathcal{U}_{\alpha s}|^{2}$ with $|\mathcal{U}_{\alpha s}|$is the active-sterile leptonic mixing matrix element which will be obtained using equation \eqref{eq:1}and $m_{s}$ represents the mass of the lightest sterile fermion.

The most important criteria for a DM candidate is its stability atleast on cosmological scale. The lightest sterile neutrino is not totally stable and may decay into an active neutrino and a photon $\gamma$ via the process $N\longrightarrow\nu+\gamma$ that leads to a monochromatic X-ray line signal. However,as discussed in many literature \cite{adhikari2017white}, the decay rate is negligible with respect to the cosmological scale because of the small mixing angle. The decay rate is given as \cite{ng2019new}:
\begin{equation}\label{eq:d}
\Gamma=1.38\times10^{-32}{\left(\frac{sin^{2}2\theta}{10^{-10}}\right)}{\left(\frac{m_{s}}{keV}\right)}^{5}s^{-1}.
\end{equation}
As seen from the above equations, the decay rate and as well as the relic abundance depend on mixing and mass of the DM candidate. Hence, the same set of model parameters which are supposed to produce correct neutrino phenomenology can also be used to evaluate the relic abundance and the decay rate of the sterile neutrino.
\section{\label{sec:level5}Constraints on keV scale sterile neutrino and effects of heavy pseudo-Dirac neutrino states}
Sterile neutrinos produced by the mixing with active neutrinos can have impacts on several phenomena. Cosmology, low energy and collider observables can severely constrain the model of keV sterile neutrinos as mentioned in \cite{abada2016impact}. Apart from this, the four pseudo-Dirac states present in the model may have significant effects on phenomena like lepton number violation, perturbativity of Yukawa couplings,unitarity violation in the active neutrino mixing matrix \cite{Chrzaszcz:2019inj},which are included in our discussion.

Firstly, the extension of standard model must comply with the latest neutrino oscillation data for both normal as well as inverted hierarchy. Our model is compatible with that as we have evaluated the model parameters using the latest global fit neutrino oscillation data \cite{de2018status} which will be discussed in the next section.

Perturbativity of Yukawa couplings is another important condition to be satisfied by any model. It puts strong upper bound on the Yukawa couplings. Our model contains three types of Yukawa couplings $y(y^{\prime})$,$y_{s}$ and $y_{r}(y_{r}^{\prime})$. The mass range of the flavon fields are so chosen that it comply with the perturbativity of Yukawa couplings in each case. Thus, in our analysis to explain neutrino phenomenology and dark matter Yukawa couplings are in the range $(10^{-5}-10^{-2})$ which satisfies the perturbativity limits. The presence of heavy pseudo-Dirac states cannot affect the perturbativity limits of Yukawa couplings. 

The addition of heavy neutrinos to the standard model may originate processes like lepton number violating interactions, among which neutrinoless double beta decay(0$\nu\beta\beta$) is the most significant one \cite{benevs2005sterile}. In our study, we have calculated the contributions of the four pseudo-Dirac states and also the sterile state to the effective electron neutrino Majorana mass $m_{ee}$ according to \cite{abada2019beta,blennow2010neutrinoless}. We have used the strong bounds on the effective mass provided by KamLAND-ZEN \cite{gando2016search} 
\begin{equation}\label{key}
	m_{ee} < 0.061- 0.165 eV
\end{equation}

In our analysis, we did not impose any unitarity violation in the active neutrino mixing matrix (PMNS matrix). In our framework, the effective light neutrino mass matrix $m_{\nu}$ can be written as
\begin{equation}
	m_{\nu} = U^{*}{m_{\nu}}U^{\dagger}
\end{equation}
The unitary matrix U is related to the PMNS mixing matrix $U^{\nu}$ as follows \cite{abada2017neutrino}
\begin{equation}
	U^{\nu} = (1-\frac{1}{2}\theta {\theta}^{\dagger})U + \mathcal{O}(\theta^{3})\simeq (1-\eta)U
\end{equation}
Here, $\theta$ can be expressed as,
\begin{equation}
	\theta = M_{d}M_{N}^{-1}
\end{equation}
In our analysis, $\eta$ lies in the range $(10^{-6}-10^{-11})$ which can be ignored in the calculations. Moreover, as mentioned by the authors in \cite{abada2017neutrino}, strong experimental constraints allow us to neglect $\theta$ which parametrises the deviation from unitarity of the PMNS matrix.

Direct detection of sterile neutrino provides a strong bounds on mass-mixing parameter space. There are several experiments for the detection of dark matter \cite{aprile2018intrinsic,campos2016testing}. We have considered the constraints from XENON100,XENON1T which have provided bounds on active-sterile mixing and our model is consistent with these bounds \cite{campos2016testing}. 

A number of cosmological observations put constraints on sum of the three neutrino masses for both normal and inverted hierarchy \cite{Aghanim:2018eyx}. It has been observed that our model is consistent with the cosmological limits on sum of the active neutrino masses.

Indirect detection is a powerful technique to search for dark matter in which involves in looking for signatures of DM decaying or annihilating into visible products \cite{boyarsky2006strategy}. Many well motivated DM candidates could produce the characteristic signature of monochromatic X ray lines. Sterile neutrinos in keV range can decay into decay into an active neutrino and a mono- energetic photon($\gamma$). This kind of signature is within reach of satellite detectors like CHANDRA \cite{horiuchi2014sterile} and XMN \cite{boyarsky2008constraints} providing significant bounds on DM mass and DM-active couplings. In our analysis, we have imposed the constraints on the mass-mixing parameter space as mentioned in \cite{abazajian2012light,perez2017almost,ng2019new}.

The study of the matter power spectrum in the intergalactic medium (IGM) has been extensively used to constrain the nature of DM \cite{Adhikari2017}. In particular,the new limits on the nature of DM comes from from the observable of the IGM in the ultraviolet and optical bands, the so called Lyman forest \cite{meiksin2009publisher}. The Lyman-$\alpha$ constraint is strongly model dependent and the bounds are related to the warm dark matter production mechanism,and to which extent this mechanism contributes to the total DM abundance \cite{abada2014dark}. Warm dark matter candidates are subjected to significant constraints from structure formation, which provides lower bounds on the DM mass \cite{boyarsky2009realistic}. In our model, sterile neutrinos are produced via non resonant production (NRP) mechanism. The Lyman-$\alpha$ data imposes strong constraint on the mass of the sterile neutrino. We have adopted the results in \cite{baur2017constraints,boyarsky2009lyman}.
\section{\label{sec:level6}Numerical Analaysis and Results}
As discussed before, the structures of the different matrices $m_{D}$, M, $\mu$ involved in ISS are formed using the discrete flavor symmetry $S_{4}$ and we obtain resulting light neutrino mass matrix using \eqref{eq:20} . The light neutrino mass matrix arising from the model is consistent with non-zero $\theta_{13}$ as $S_{4}$ product rules lead to the light neutrino mass matrix in which the $\mu-\tau$ symmetry is explicitly broken. The neutrino mass matrix contains eight complex model parameters. For the numerical analysis,we have fixed the value of two model parameters. The remaining six parameters can be evaluated by comparing the neutrino mass matrix arising from the model with the one which is parametrized by the available $3\sigma$ global fit data given in table \ref{tab3}. This parametric form of light neutrino mass matrix is complex symmetric and so it contains six independent complex elements. Thus we can solve the six model parameters in terms of the known neutrino parameters using equation \eqref{eq:16}. We first randomly generate the nine light neutrino parameters in the $3\sigma$ range and then we evaluate the six model parameters. The light neutrino mass matrix can be written as,
\begin{equation}\label{eq:16}
m_{\nu} = U_{\text{PMNS}}m^{\text{diag}}_{\nu} U^T_{\text{PMNS}}
\end{equation}
where the Pontecorvo-Maki-Nakagawa-Sakata (PMNS) leptonic mixing matrix can be parametrized as \cite{giganti2017neutrino}
\begin{equation}
U_{\text{PMNS}}=\left(\begin{array}{ccc}
c_{12}c_{13}& s_{12}c_{13}& s_{13}e^{-i\delta}\\
-s_{12}c_{23}-c_{12}s_{23}s_{13}e^{i\delta}& c_{12}c_{23}-s_{12}s_{23}s_{13}e^{i\delta} & s_{23}c_{13} \\
s_{12}s_{23}-c_{12}c_{23}s_{13}e^{i\delta} & -c_{12}s_{23}-s_{12}c_{23}s_{13}e^{i\delta}& c_{23}c_{13}
\end{array}\right) U_{\text{Maj}}
\label{matrixPMNS}
\end{equation}
where $c_{ij} = \cos{\theta_{ij}}, \; s_{ij} = \sin{\theta_{ij}}$ and $\delta$ is the leptonic Dirac CP phase. The diagonal matrix $U_{\text{Maj}}=\text{diag}(1, e^{i\alpha}, e^{i(\beta+\delta)})$  contains the Majorana CP phases $\alpha, \beta$.
The diagonal mass matrix of the light neutrinos can be written  as, $m^{\text{diag}}_{\nu}
= \text{diag}(0, \sqrt{m^2_1+\Delta m_{solar}^2}, \sqrt{m_1^2+\Delta m_{atm}^2})$ for normal hierarchy and  $m^{\text{diag}}_{\nu} = \text{diag}(\sqrt{m_3^2+\Delta m_{atm}^2}, 
\sqrt{\Delta m_{solar}^2+ \Delta m_{atm}^2}, m_3)$ for inverted hierarchy \cite{Nath_2017}.
\begin{table}
	\centering
	\begin{tabular}{|c|c|c|}
		
		\hline 
		Oscillation parameters	& 3$\sigma$(NO) & 3$\sigma$(IO) \\ 
		\hline 
		$\frac{\Delta m_{21}^{2}}{10^{-5}eV^{2}}$	& 7.05 - 8.14 &7.05 - 8.14  \\ 
		\hline 
		$\frac{\Delta m_{31}^{2}}{10^{-3}eV^{2}}$	&  2.41 - 2.60  &  2.31-2.51  \\ 
		\hline 
		$sin^{2}\theta_{12}$ &0.273 - 0.379  & 0.273 - 0.379 \\ 
		\hline 
		$sin^{2}\theta_{23}$ &  0.445 - 0.599  &  0.453 - 0.598\\ 
		\hline 
		$sin^{2}\theta_{13}$ &  0.0196 - 0.0241 &  0.0199 - 0.0244 \\ 
		\hline 
		$\frac{\delta}{\pi}$ & 0.87 - 1.94 &  1.12- 1.94\\ 
		\hline 
	\end{tabular} 
	\caption{Latest Global fit neutrino oscillation Data \cite{de2018status}.}\label{tab3}
\end{table} 
For numerical analysis, we have fixed the value of $f= 2\times10^{10}$eV and $g= 10^{12}$eV and calculated the other model parameters a,b,c,e,h in our model. The fixing of model parameters is done considering the range given in \cite{abada2017neutrino}to get the desired DM mass and mixing. Author \cite{abada2017neutrino} , specified that the first row of the submatrix $M_{N}$ determines the mass scale of the lightest pseudo-Dirac pair and the second row of the submatrix $M_{N}$ determines the mass scale of the heavier pseudo-Dirac pair. In our model, parameters f and h give the range of the lightest pseudo-Dirac pair and g corresponds to the heavier one. Keeping this in mind, we have fixed two model parameters from the given range.
The eigenvalues of $M_{H}$ e.g.in case of model are obtained as 
\begin{equation}
M_{H1} = \frac{1}{2}(p - \sqrt{2f^{2}+2g^{2}+2h^{2}-2\sqrt{-4f^{2}g^{2}+(-f^{2}-g^{2}-h^{2})^{2}}+p^{2}}
\end{equation}
\begin{equation}
M_{H2} = \frac{1}{2}(p+ \sqrt{2f^{2}+2g^{2}+2h^{2}-2\sqrt{-4f^{2}g^{2}+(-f^{2}-g^{2}-h^{2})^{2}}+p^{2}}
\end{equation}
\begin{equation}
M_{H3} =\frac{1}{2}(p - \sqrt{2f^{2}+2g^{2}+2h^{2}+2\sqrt{-4f^{2}g^{2}+(-f^{2}-g^{2}-h^{2})^{2}}+p^{2}}
\end{equation}
\begin{equation}
M_{H4} = \frac{1}{2}(p + \sqrt{2f^{2}+2g^{2}+2h^{2}+2\sqrt{-4f^{2}g^{2}+(-f^{2}-g^{2}-h^{2})^{2}}+p^{2}}
\end{equation}. 
\begin{equation}
M_{H5} = p
\end{equation}
It is evident that one of the eigenvalues depends only on the mass matrix $\mu$ and hence may be considered as the lightest sterile state as discussed in the previous section. This eigenvalue corresponds to the mass of the Dark Matter $m_{DM}$. Thus the model parameter p will determine the mass of the DM. The DM-active mixing can be determined from  \eqref{eq:2} and we plot the mixing as a function of model parameters as well as the DM mass. One can also obtain the mass of the dark matter from the eigenvalues of the $8\times8$ matrix M in equation \eqref{eq:2}. In our model, three eigenvalues correspond to the light active neutrino masses, four eigenvalues in TeV range give the mass of the pseudo-Dirac particles while one of the eigenvalues lie in keV range which is considered as DM mass. To get the active-sterile mixing, first we numerically diagonalise the matrix M in equation \eqref{eq:1} using the evaluated model parameters to obtain $\mathcal{U}$ and then we find the first three components of the eigenvector corresponding to the eigenvalue $m_{DM}$ which represent the three active-sterile mixing elements.

Using the same set of model parameters,numerically evaluated for the model, we have also calculated the relic abundance using equation \eqref{eq:c} and plot it as a function of the Dark Matter mass.
The decay rates of the sterile DM are obtained using\eqref{eq:d} which have been plotted as a function of the Dark Matter mass as it imposes powerful constraint on DM mass.

The masses of the heavy neutrinos and their mixing with the active neutrinos are subjected to various experimental constraints like neutrinoless double beta decay (0$\nu\beta\beta$). The decay width is proportional to the effective electron neutrino majorana mass ($m_{ee}$) and the standard contribution is given as,
\begin{equation}
m_{ee} = \mathrel{\Big|}\sum_{i = 1}^{3}{U_{ei}}^{2}m_{i}\mathrel{\Big|}
\end{equation}
Addition of extra fermions may contribute to the effective electron neutrino Majorana mass \cite{Abada:2017jjx} We have evaluated the contribution to ($m_{ee}$) using the equation in \cite{Awasthi:2013we,abada2019beta} 
\begin{equation}
M_{ee} =\mathrel{\Big|}\sum_{i = 1}^{3}{U_{ei}}^{2}m_{i}\mathrel{\Big|} + \mathrel{\Big|}\sum_{j = 1}^{5}{U_{ej}}^{2}\frac{M_{j}}{k^{2}+M_{j}^{2}}|<k>|^{2}\mathrel{\Big|}
\end{equation}
where,  $M_{j}$ and ${U_{ej}}$ represent the mass of the heavy neutrinos and couplings to the electron neutrino respectively. $|<k>|\simeq 190$ MeV represents neutrino virtuality momentum. The mixing of the heavy neutrinos to the electron neutrino are obtained from the mixing matrix $\mathcal{U}$. For this, the four eigenvectors corresponding to four TeV ranged eigenvalues are evaluated and then the first components of four eigenvectors give the mixing of the heavy neutrinos to the electron neutrino.

Thus using the same set of model parameters that are used to give correct neutrino phenomenology, we have calculated the the effective electron neutrino majorana mass in (0$\nu\beta\beta$), DM-Active mixing, decay rate of the lightest sterile neutrino and relic abundance for normal hierarchy(NH) as well as inverted hierarchy(IH).

\begin{figure}
	\begin{center}
		\includegraphics[width=0.45\textwidth]{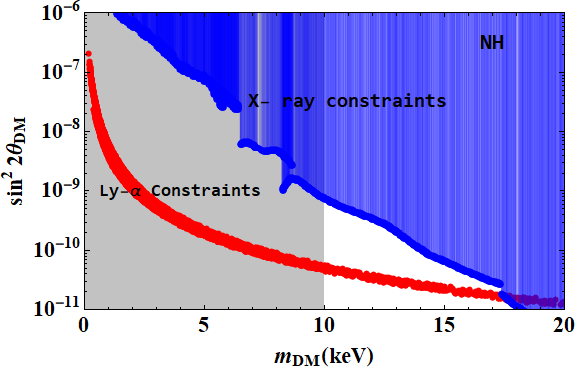}
		\includegraphics[width=0.45\textwidth]{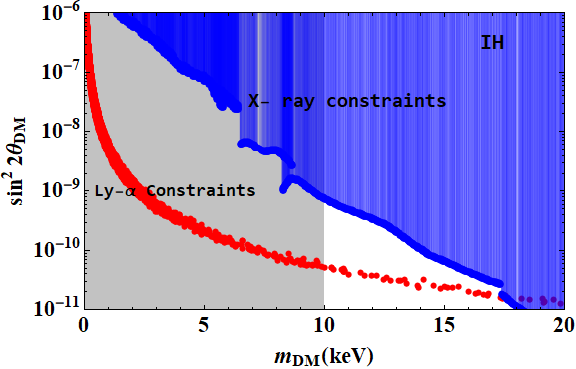}
	\end{center}
	\begin{center}
		\caption{DM-active mixing as a function of the mass of the DM for both normal hierarchy as well as inverted hierarchy.}
		\label{fig1}
	\end{center}
\end{figure}

\begin{figure*}
	\begin{center}
	    \includegraphics[width=0.45\textwidth]{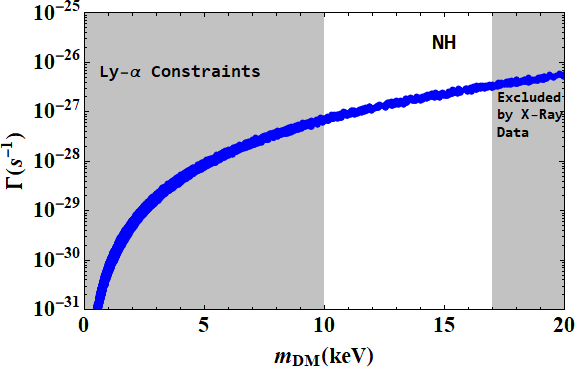}		\includegraphics[width=0.45\textwidth]{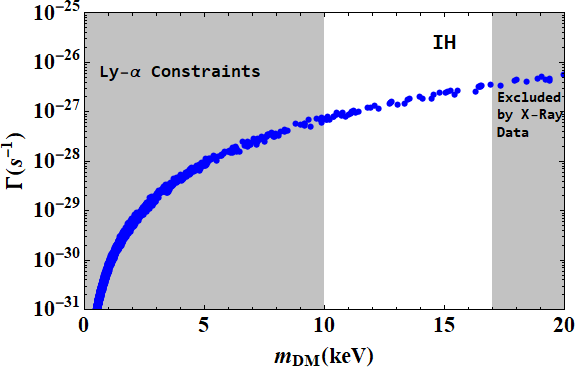}
	\end{center}
	\begin{center}
		\caption{Decay rate of the lightest sterile neutrino as a function of DM mass for normal hierarchy as well as inverted hierarchy.}
		\label{fig2}
	\end{center}
\end{figure*}

\begin{figure*}
	\begin{center}
		\includegraphics[width=0.45\textwidth]{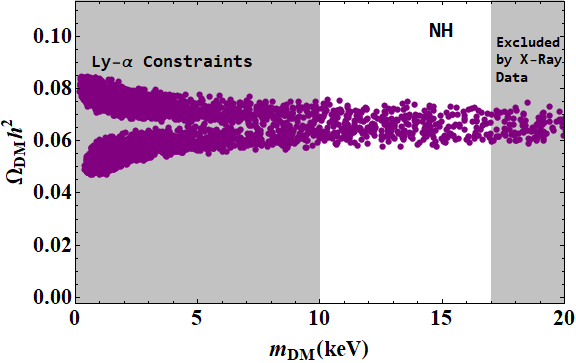}
		\includegraphics[width=0.45\textwidth]{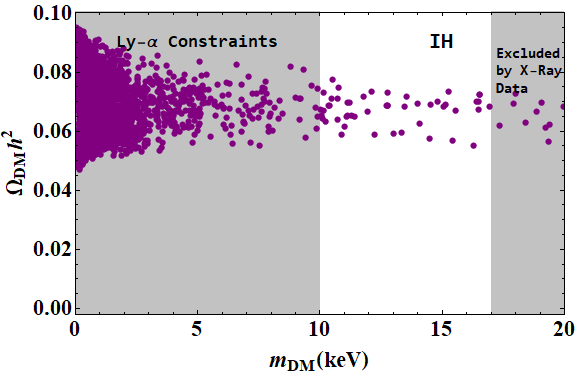}
	\end{center}
	\begin{center}
		\caption{Sterile neutrino contribution to DM abundance in NH and IH}
		\label{fig3}
	\end{center}
\end{figure*}

\begin{figure*}
	\begin{center}
		\includegraphics[width=0.40\textwidth]{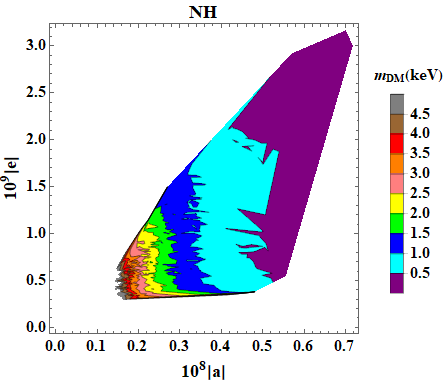}
		\includegraphics[width=0.40\textwidth]{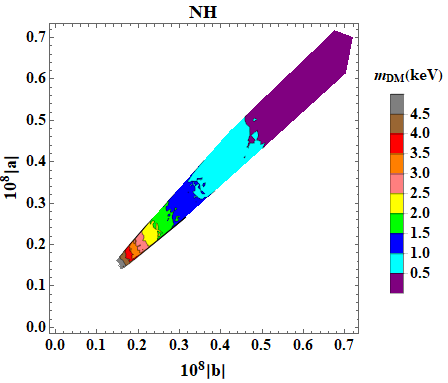}\\ 
		\includegraphics[width=0.40\textwidth]{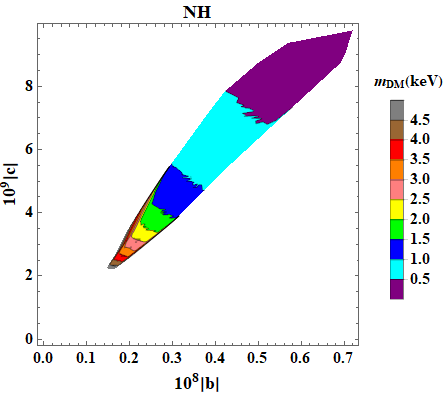}
		\includegraphics[width=0.40\textwidth]{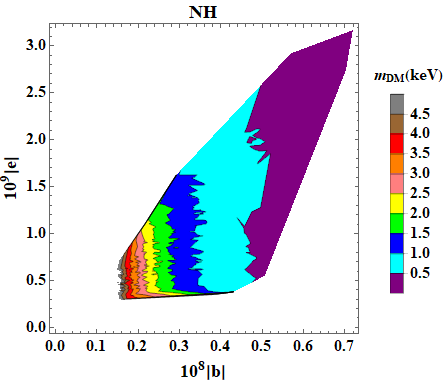}
	\end{center}
	\begin{center}
		\caption{Variation of different model parameters (in eV) with Dark Matter mass as a contour for normal hierarchy. The mass limit as predicted by cosmology is $(0.4-50)$keV.}
		\label{fig4}
	\end{center}
\end{figure*}

\begin{figure*}
	\begin{center}		
		\includegraphics[width=0.40\textwidth]{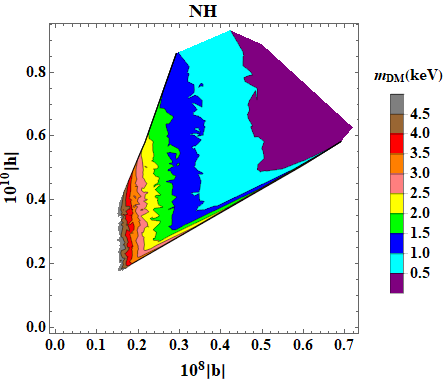}
		\includegraphics[width=0.40\textwidth]{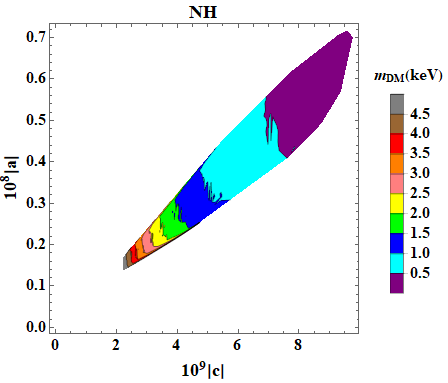}\\
		\includegraphics[width=0.40\textwidth]{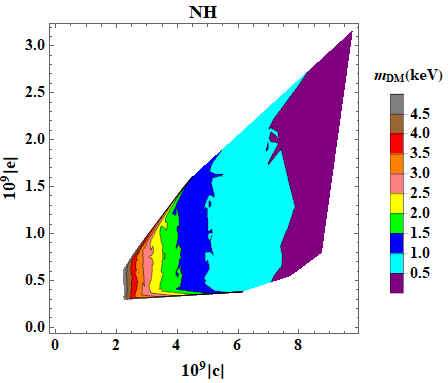}
		\includegraphics[width=0.40\textwidth]{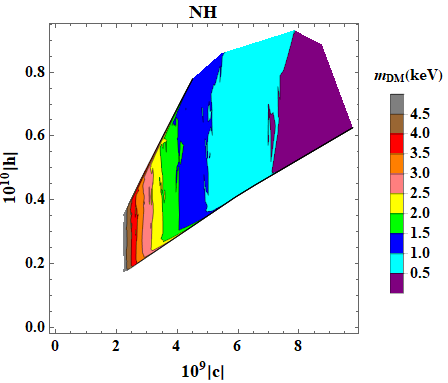}
	\end{center}
	\begin{center}
		\caption{Variation of different model parameters (in eV)  with Dark Matter mass as a contour for normal hierarchy. The mass limit as predicted by cosmology is $(0.4-50)$keV.}
		\label{fig5}
	\end{center}
\end{figure*}

\begin{figure*}
	\begin{center}
		\includegraphics[width=0.40\textwidth]{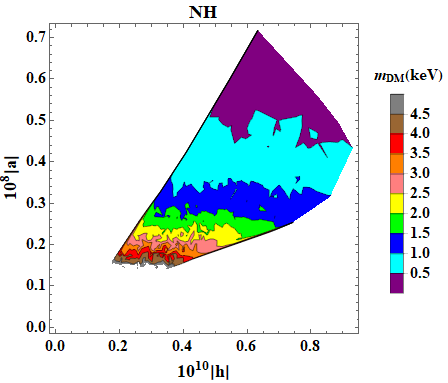}
		\includegraphics[width=0.40\textwidth]{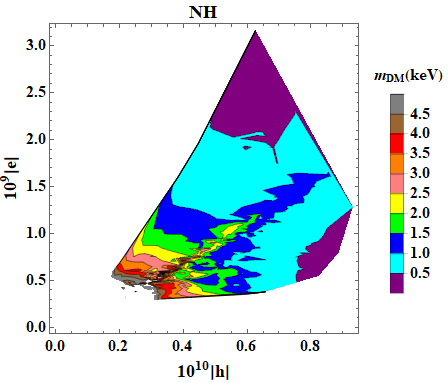}
	\end{center}	
	\begin{center}
		\caption{Variation of different model parameters (in eV) with Dark Matter mass as a contour for normal hierarchy. The mass limit as predicted by cosmology is $(0.4-50)$keV.}
		\label{fig6}
	\end{center}
\end{figure*}

\begin{figure*}
	\begin{center}
		\includegraphics[width=0.40\textwidth]{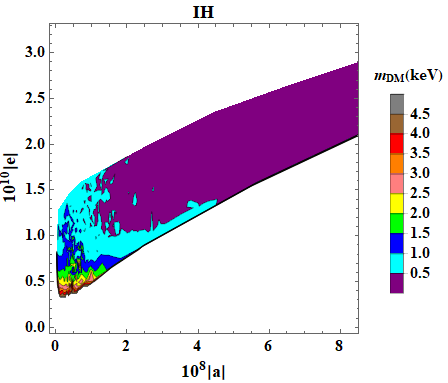}
		\includegraphics[width=0.40\textwidth]{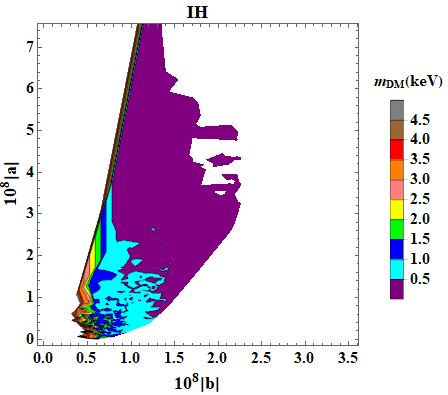}\\ 
		\includegraphics[width=0.40\textwidth]{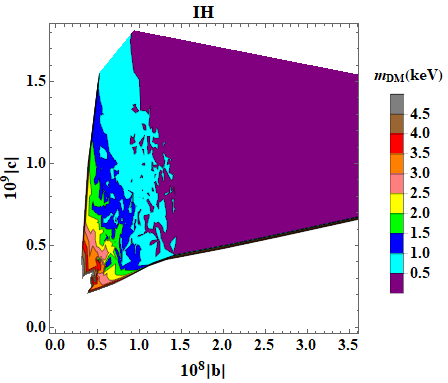}
		\includegraphics[width=0.40\textwidth]{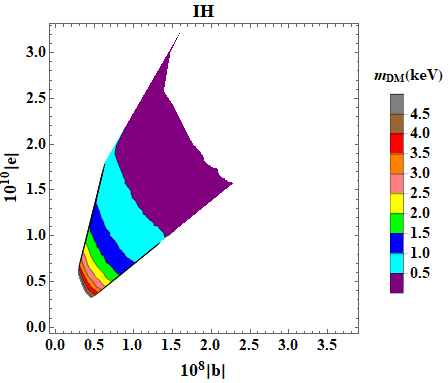}
	\end{center}	
	\begin{center}
	\caption{Variation of different model parameters (in eV) with Dark Matter mass as a contour for inverted hierarchy The mass limit as predicted by cosmology is $(0.4-50)$keV.}
		\label{fig7}
	\end{center}
\end{figure*}

\begin{figure*}
	\begin{center}	
		\includegraphics[width=0.40\textwidth]{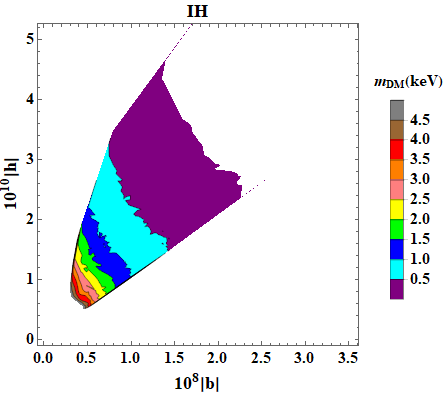}
		\includegraphics[width=0.40\textwidth]{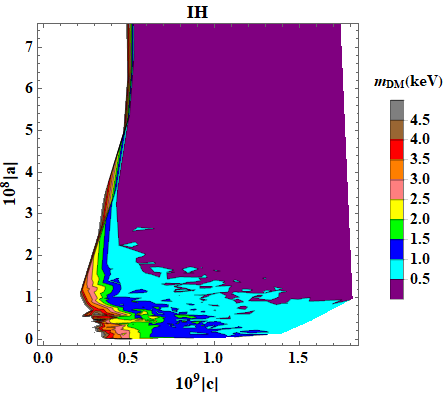}\\
		\includegraphics[width=0.40\textwidth]{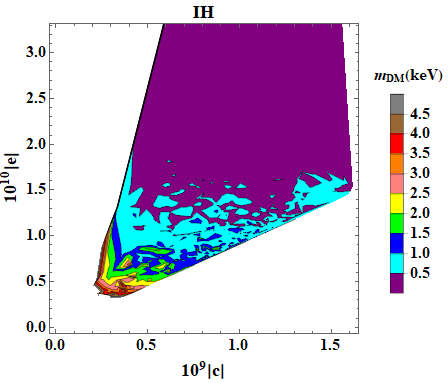}
		\includegraphics[width=0.40\textwidth]{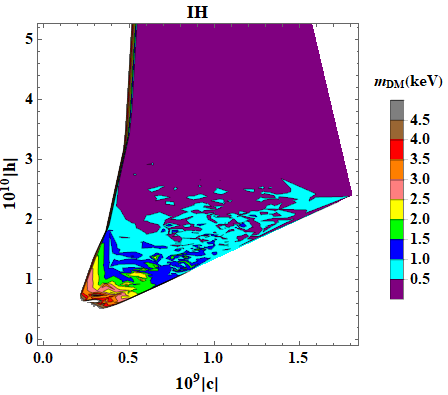}
	\end{center}	
	\begin{center}
	\caption{Variation of different model parameters (in eV) with Dark Matter mass as a contour for inverted hierarchy The mass limit as predicted by cosmology is $(0.4-50)$keV.}
		\label{fig8}
	\end{center}
\end{figure*}

\begin{figure*}
	\begin{center}		
		\includegraphics[width=0.40\textwidth]{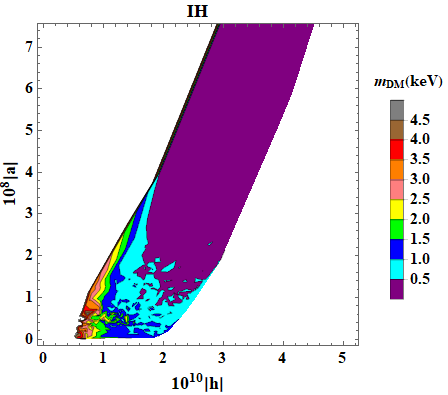}
		\includegraphics[width=0.40\textwidth]{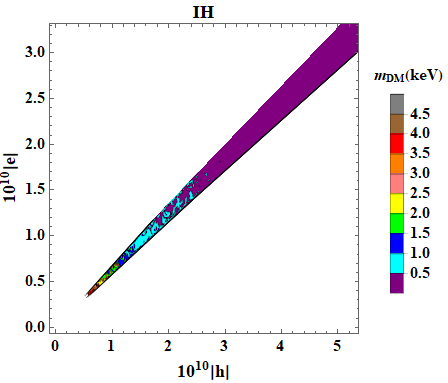}
	\end{center}	
	\begin{center}
		\caption{Variation of different model parameters (in eV) with Dark Matter mass as a contour for inverted hierarchy The mass limit as predicted by cosmology is $(0.4-50)$keV.}
		\label{fig9}
	\end{center}
\end{figure*}

\begin{figure*}
	\begin{center}
		\includegraphics[width=0.40\textwidth]{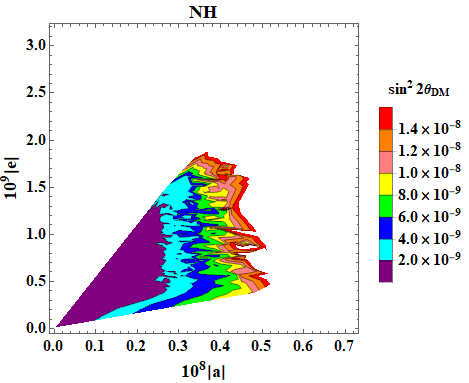}
		\includegraphics[width=0.40\textwidth]{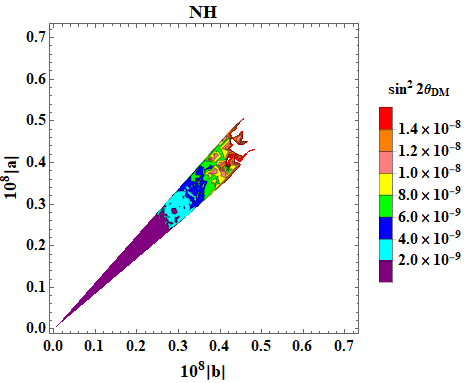}\\ 
		\includegraphics[width=0.40\textwidth]{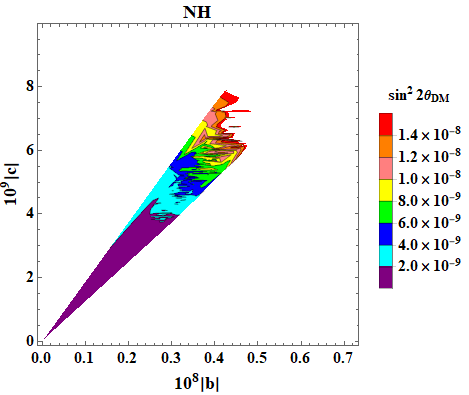}
		\includegraphics[width=0.40\textwidth]{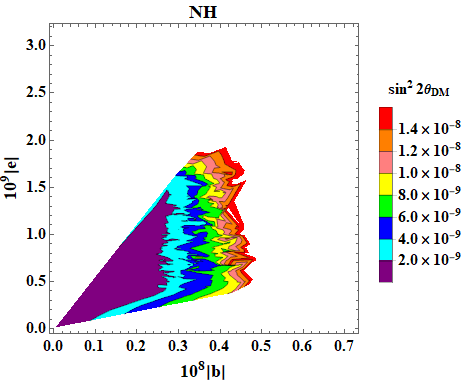}
	\end{center}
	\begin{center}
		\caption{Variation of different model parameters (in eV) with the DM-active mixing as the contour for normal hierarchy. The limit as predicted by cosmology is $(10^{-12}-10^{-8})$}
		\label{fig10}
	\end{center}
\end{figure*}

\begin{figure*}
	\begin{center}		
		\includegraphics[width=0.40\textwidth]{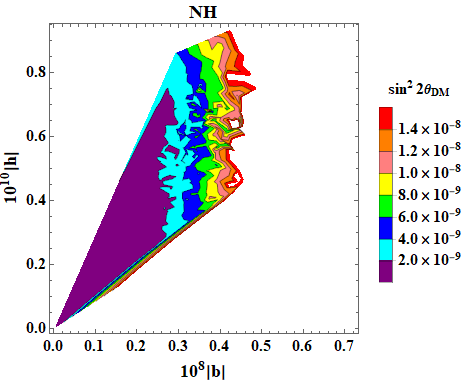}
		\includegraphics[width=0.40\textwidth]{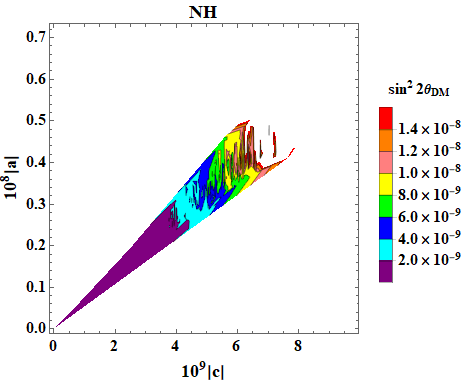}\\
		\includegraphics[width=0.40\textwidth]{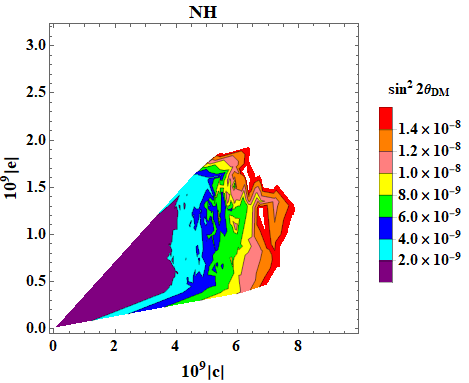}
		\includegraphics[width=0.40\textwidth]{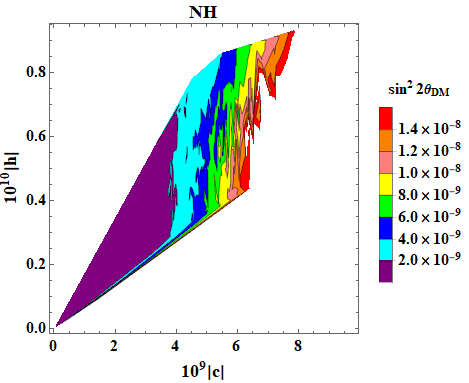}
	\end{center}	
	\begin{center}
		\caption{Variation of different model parameters (in eV) with the DM-active mixing as the contour for normal hierarchy. The limit as predicted by cosmology is ($10^{-12}-10^{-8})$}
		\label{fig11}
	\end{center}
\end{figure*}

\begin{figure*}
	\begin{center}	
		\includegraphics[width=0.40\textwidth]{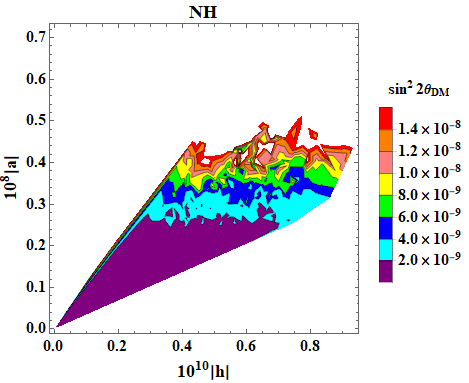}
		\includegraphics[width=0.40\textwidth]{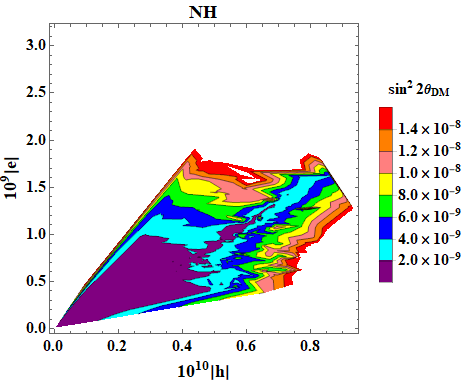}
	\end{center}	
	\begin{center}
		\caption{Variation of different model parameters (in eV) with the DM-active mixing as the contour for normal hierarchy. The limit as predicted by cosmology is $(10^{-12}-10^{-8})$}
		\label{fig12}
	\end{center}
\end{figure*}

\begin{figure*}
	\begin{center}
		\includegraphics[width=0.40\textwidth]{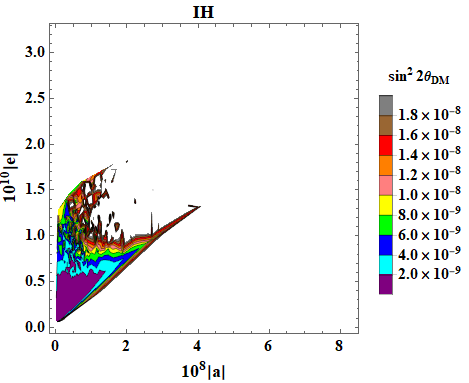}
		\includegraphics[width=0.40\textwidth]{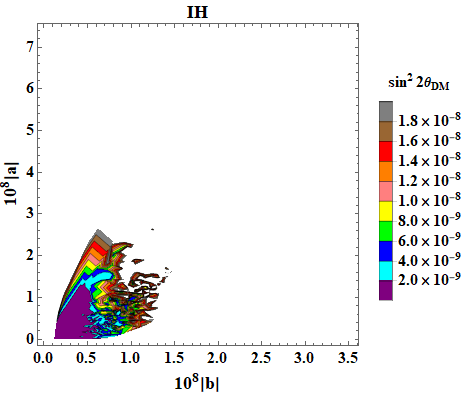}\\ 
		\includegraphics[width=0.40\textwidth]{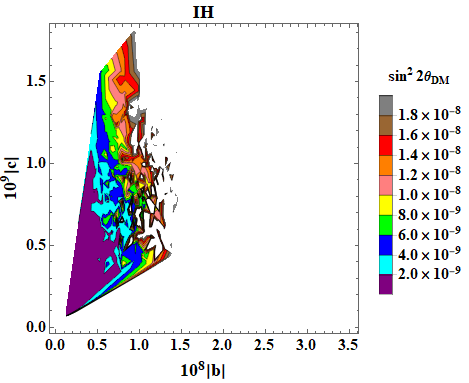}
		\includegraphics[width=0.40\textwidth]{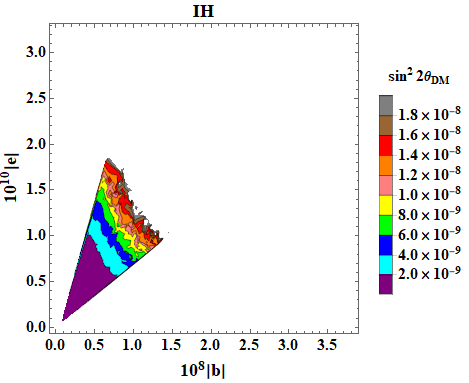}
	\end{center}	
	\begin{center}
		\caption{Variation of different model parameters (in eV) with the DM-active mixing as the contour for inverted hierarchy. The limit as predicted by cosmology is  $(10^{-12}-10^{-8})$}
		\label{fig13}
	\end{center}
\end{figure*}

\begin{figure*}
	\begin{center}		
		\includegraphics[width=0.40\textwidth]{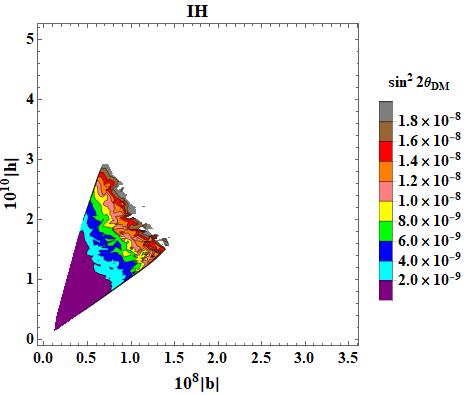}
		\includegraphics[width=0.40\textwidth]{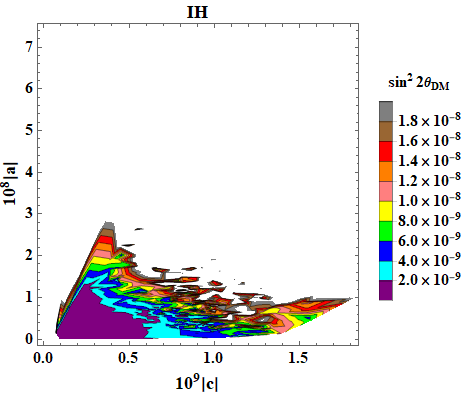}\\
		\includegraphics[width=0.40\textwidth]{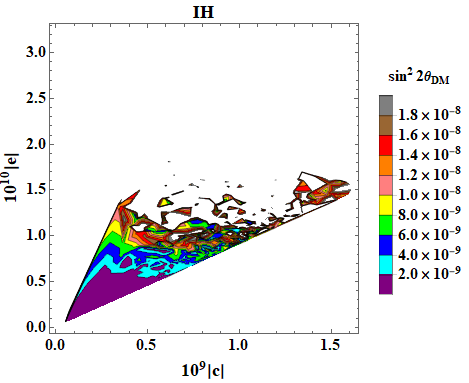}
		\includegraphics[width=0.40\textwidth]{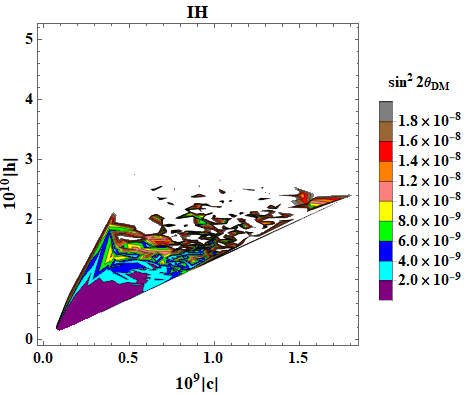}
	\end{center}	
	\begin{center}
		\caption{Variation of different model parameters (in eV) with the DM-active mixing as the contour for inverted hierarchy. The limit as predicted by cosmology is  $(10^{-12}-10^{-8})$}
		\label{fig14}
	\end{center}
\end{figure*}

\begin{figure*}
	\begin{center}		
		\includegraphics[width=0.40\textwidth]{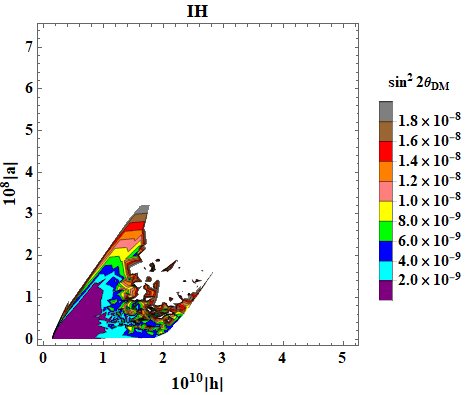}
		\includegraphics[width=0.40\textwidth]{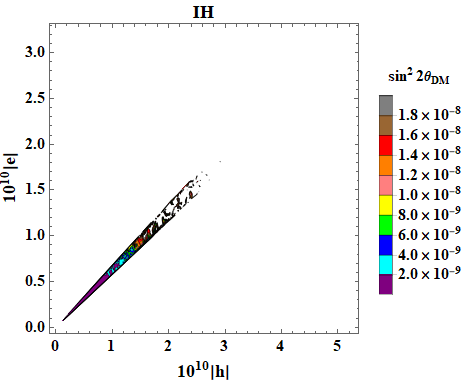}
	\end{center}	
	\begin{center}
		\caption{Variation of different model parameters (in eV) with the DM-active mixing as the contour for inverted hierarchy. The limit as predicted by cosmology is  $(10^{-12}-10^{-8})$}
		\label{fig15}
	\end{center}
\end{figure*}

\begin{figure}
	\begin{center}
		\includegraphics[width=0.45\textwidth]{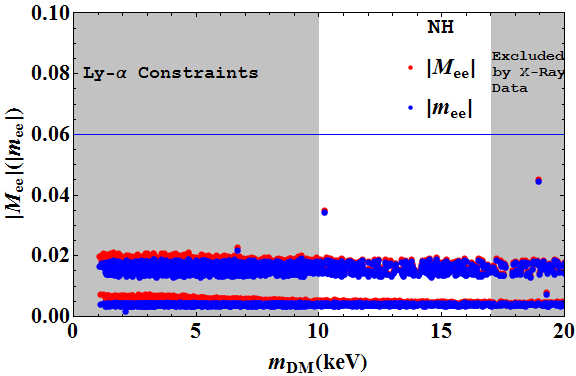}
		\includegraphics[width=0.45\textwidth]{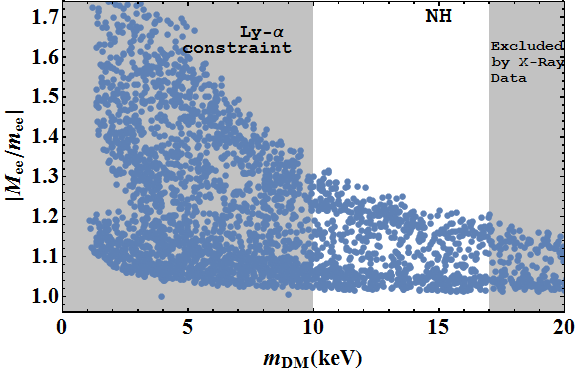}
	\end{center}
	\begin{center}
		\caption{Preictoions of effective mass as a function of $m_{DM}$ for normal hierarchy. The standard contribution is shown in blue color and
			nonstandard contributions with pseudo-Dirac neutrinos and sterile state are shown in red color. Comparison of these two contributions are shown in the figure on the right side.}
		\label{fig16}
	\end{center}
\end{figure}

\begin{figure}
	\begin{center}
		\includegraphics[width=0.45\textwidth]{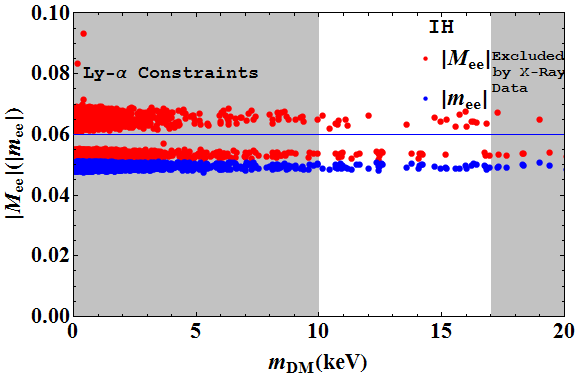}
		\includegraphics[width=0.45\textwidth]{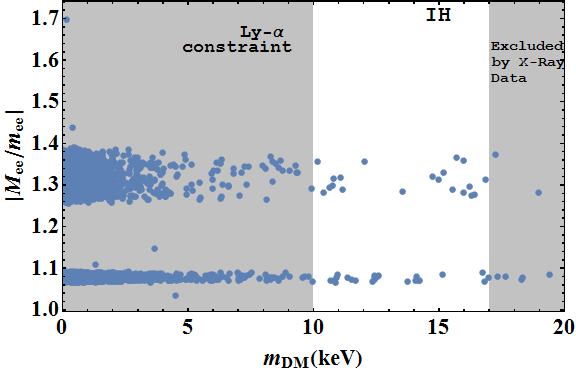}
	\end{center}
	\begin{center}
		\caption{Effective mass as a function of $m_{DM}$ for inverted hierarchy.The standard contribution is shown in blue color and
			nonstandard contributions with pseudo-Dirac neutrinos and sterile state are shown in red color. Comparison of these two contributions are shown in the figure on the right side.}
		\label{fig17}
	\end{center}
\end{figure}

\begin{figure}
	\begin{center}
		\includegraphics[width=0.45\textwidth]{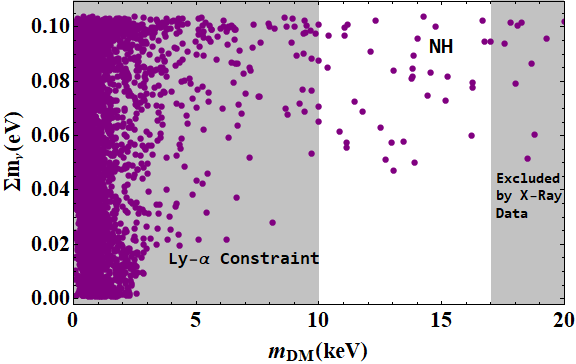}
		\includegraphics[width=0.45\textwidth]{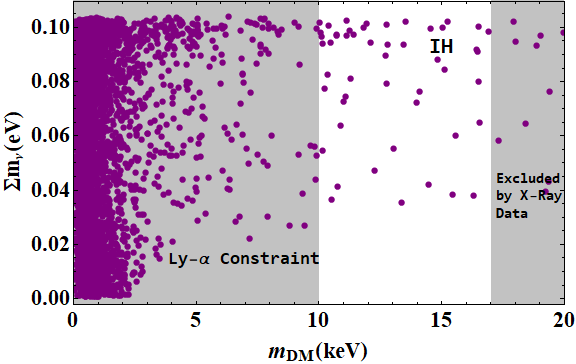}
	\end{center}
	\begin{center}
		\caption{Sum of light neutrino masses as a function of mass of DM for both normal hierarchy as well as inverted hierarchy.}
		\label{fig18}
	\end{center}
\end{figure}

\begin{itemize}
	\item {We have shown different plots obtained from our numerical analysis carried out for normal hierarchy as well as inverted hierarchy from fig \ref{fig1} to fig \ref{fig17}.}
	\item {Fig \ref{fig1} indicates the parameter space for DM mass and DM-active mixing for normal and inverted hierarchy. To be a good DM candidate, sterile neutrino must have mass in the range (0.4-50)keV and should mix with active neutrinos within the range($10^{-12}-10^{-8}$). From the figure it is clear that both the mass and mixing satisfy the cosmological limit. We have incorporated the cosmological bounds from Lyman-$\alpha$ and X-ray data in the figure.}
\item {Fig \ref{fig2} shows the prediction of decay rate of the lightest sterile neutrino as a function of DM mass for both the mass hierarchies. It is seen that the decay rate is very negligible and in the range $10^{-31}$ to  $10^{-27}$ $s^{-1}$  for both the hierarchies. The decay rate imposes influential constraint on DM mass and the constraints from structure formation are imposed in the figure. From fig \ref{fig1},  it is evident that X-ray limits exclude the DM masses above 17 keV which has also been incorporated in the fig \ref{fig2}}
\item {We have shown the relic abundance of the proposed DM candidate as a function of DM mass in fig \ref{fig3} . The relic abundance shown here indicates the partial contribution of sterile neutrinos to the total DM abundance. Our model can account for almost $66\%$ of the total DM abundance for NH and almost $83\%$ of the total DM abundance for IH. The limits on the mass of the dark matter from Lyman-$\alpha$ and X-ray data as obtained from the previous figures are also imposed in fig \ref{fig3}. The allowed region of the parameter space is spanned from 10 keV to 17 keV.}
\item {Fig \ref{fig4}, \ref{fig5} and \ref{fig6} show the two parameter contour plots with DM mass as contour in NH and Fig \ref{fig7}, \ref{fig8} and \ref{fig9} show the two parameter contour plots with DM mass as contour in IH. In our model, we have six model parameters and one of them corresponds to DM mass. Hence, we are left with five model parameters and 10 different combinations of parameters which are shown. It is observed that for both the hierarchies, the model parameters give rise to DM mass within the cosmological limit. The values of the model parameters giving rise to the allowed DM mass are given in table \ref{tab5}.}
\item {Fig \ref{fig10}, \ref{fig11} and \ref{fig12} show the two parameter contour plots with DM-active mixing as contour in NH while fig \ref{fig13}, \ref{fig14} and \ref{fig15} represent  the two parameter contour plots with DM-active mixing as contour in IH. Here also, we have 10 different combinations of model parameters. It is observed that for both the hierarchies, the model parameters give rise to DM-active mixing within the range predicted by cosmology. The allowed values of the model parameters are shown in table \ref{tab6}.}
\item {Fig \ref{fig16} and Fig \ref{fig17} shows the effects of the heavy states present in the model on effective mass as a function of $m_{DM}$. Here, $m_{ee}$ represents the standard contribution while $M_{ee}$ is the contribution from the pseudo-Dirac states as well as the sterile state. The presence of pseudo-Dirac states along with the sterile state increases the effective mass parameter upto 1.7 times (1.4 times) larger than the standard contribution for NH (IH). It is observed that the change is significant below 10 keV of dark matter mass which range is however excluded by $Ly-\alpha$ data. It is evident from the figures that the model parameters give rise to effective mass within the experimental limits provided by KamLAND-ZEN for both the hierarchies.}
	\item {Fig \ref{fig18} indicates the variation of sum of the light neutrino masses as a function of dark matter mass $m_{DM}$ for NH and IH. It is observed that for both the hierarchies, our model is consistent with the cosmological limits on the sum of the light neutrino masses.}
\end{itemize}
\begin{table}
	\centering
	\begin{tabular}{|c|c|c|}
		\hline 
		Model Parameter  & NH(eV) & IH(eV) \\ 
		\hline 
		a & $1.5\times 10^{7}-7.5\times 10^{7}$ &  $ 1\times 10^{7}-8.0\times 10^{8}$\\
		\hline
		b & $1.5\times 10^{7}-7.7\times 10^{7}$ &  $4.0\times 10^{7}-2.5\times 10^{8}$\\
		\hline 
		c & $2.0\times 10^{9}-1.0\times10^{10}$ &  $2.5\times 10^{8}-1.7\times 10^{9}$ \\
		\hline 
		e & $2.5\times 10^{8}-3.4\times 10^{9}$ &  $2.5\times 10^{9}-2.7\times 10^{10}$\\
		\hline
		h & $2.0\times 10^{9}- 1.0\times10^{10}$ &  $0.5\times 10^{10}-4.8\times 10^{10}$\\
		\hline
	\end{tabular} 
	\caption{Allowed range of the model parameters satisfying DM mass.} \label{tab5}
\end{table} 
\begin{table}
	\centering
	\begin{tabular}{|c|c|c|}
		\hline 
		Model Parameter  & NH(eV) & IH(eV) \\ 
		\hline 
		a & $0.1\times 10^{7}-5.0\times 10^{7}$ & $0.1\times 10^{7}-3.0\times 10^{8}$\\
		\hline
		b & $0.1\times 10^{7}-5.0\times 10^{7}$ &  $1.0\times 10^{7}-1.5\times 10^{8}$\\
		\hline 
		c & $0.1\times 10^{9}-8.0\times 10^{9}$ &  $1.0\times 10^{8}-1.9\times 10^{9}$ \\
		\hline 
		e & $0.1\times 10^{8}-2.0\times 10^{9}$ &  $1.0\times 10^{9}-1.7\times 10^{10}$\\
		\hline
		h & $0.1\times 10^{9}- 9.5\times 10^{9}$ &  $2.0\times 10^{9}-3.0\times 10^{10}$\\
		\hline
	\end{tabular} 
	\caption{Allowed range of the model parameters satisfying DM-active mixing.} \label{tab6}
\end{table}
\section{\label{sec:level7}Discussion and Conclusion}
Searching for the particle nature of dark matter is a challenging issue to the physics community worldwide. Motivated by this, we have conducted our study in the framework of ISS(2,3) which leads to three light active neutrino states, two pseudo-Dirac states and an additional sterile state in keV range which can be considered as a dark matter candidate. The motivation of the work is to search for a common framework which can simultaneously address the dark matter problem as well as neutrino phenomenology. We have considered the model which is the extension of standard model by discrete flavor symmetry  $S_{4}\times Z_{4} \times Z_{3}$ that can simultaneously explain the correct neutrino oscillation data and also provides a good dark matter candidate. The different mass matrices involved in ISS(2,3) are constructed using $S_{4}$ product rules which lead to a light neutrino mass matrix $ m_{\nu}$ which is consistent with the broken $ \mu-\tau $ symmetry.

We have evaluated the model parameters by comparing the light neutrino mass matrix with the one constructed from light neutrino parameters for both the hierarchical pattern of neutrino mass spectrum(NH and IH). After finding the model parameters, we then use the allowed parameters to calculate the mass of the lightest sterile neutrino which is considered as dark matter candidate and also the DM-active mixing. We obtain two parameter contour plots taking dark matter mass and DM-active mixing as contours. It is quite interesting that a wide range of model parameters give dark matter mass and DM-active mixing within the cosmological range for both normal as well as inverted hierarchy. By comparing different plots for different combinations of the model parameters, we have found a common allowed parameter space considering the requirement of satisfying the correct DM mass and mixing within the model framework. Considering the lightest sterile neutrino as dark matter candidate, we have calculated the relic abundance which accounts for almost $\sim$ 83 $\%$ of the total DM abundance. We have also verified the stability of the dark matter by calculating the decay rate of the lightest sterile neutrino in the process $N\longrightarrow\nu+\gamma$ (N being the sterile neutrino)  with same range of allowed parameters. We have obtained a negligible decay rate which ensures the stability of the dark matter candidate at least in the cosmological scales. We have also verified the viability of the model from different phenomena like constraints from direct searches of sterile neutrinos, deviation from unitarity of the PMNS mixing matrix, perturbativity of the Yukawa couplings, neutrinoless double beta decay. 
Thus we present a model which offers a well-motivated DM candidate as well as correct neutrino phenomenology. The detailed numerical analysis assures sufficient parameter space satisfying the criteria of a feasible dark matter candidate and a better reach for neutrino sector as well. A wide range of parameter space allowed from the dark matter consideration, has the possibility to be probed at LHC in near future. The keV scale sterile neutrino can play a vital role in successfully achieving low scale leptogenesis, which we leave for our future study. 
\appendix
\section{Properties of $S_4$ group} 
\label{appen1}

$S_{4}$ is the group of permutations of four objects with 24 group elements.
S and T are the two generators \cite{ishimori2012introduction} of $S_{4}$ satisfying
\begin{equation} \label{A1}
T^{4} = S^{3} = e ,   TS^{2}T = S
\end{equation}
All of the $S_{4}$ elements can be written as products of these two generators. There are five inequivalent irreducible representations of $S_{4}$, two singlets 1 and $1^{\prime}$, one doublet 2 and 
two triplets 3 and $3^{\prime}$. S and T have different structures depending on which irreducible representation we are considering singlet, doublet or triplet. 
The representations are given as follows
\begin{gather*}
a,b \sim1_{1} ,
\left(\begin{array}{c}
a_{1}\\
a_{2}\end{array}\right) ,  
\left(\begin{array}{c}
b_{1}\\
b_{2}\end{array}\right) \sim 2 , \left(\begin{array}{c}
a_{1}\\
a_{2}\\
a_{3}
\end{array}\right),  \left(\begin{array}{c}
b_{1} \\
b_{2} \\
b_{3}\end{array}\right) \sim 3 , \left(\begin{array}{c}
a_{1}^{\prime} \\
a_{2}^{\prime} \\
a_{3}^{\prime}
\end{array}\right),  \left(\begin{array}{c}
b_{1}^{\prime} \\
b_{2}^{\prime} \\
b_{3}^{\prime} \end{array}\right) \sim 3^{\prime}.
\end{gather*} 

The tensor products of $S_{4}$ that has been used in the present analysis are given below (for more details see \cite{ishimori2012introduction})
\begin{equation*}
3 \times 1 = 3, 3 \times 1^{\prime} = 3^{\prime}, 3^{\prime} \times 1^{\prime} = 3 ,  2 \times 1^{\prime}=2.\\
\end{equation*}

\begin{gather}
(A)_{3} \times (B)_{3} = \left(A \cdot B \right)_{1} + \left(\begin{array}{c}
A \cdot \Sigma \cdot B\\
A \cdot \Sigma^{*} \cdot B \end{array}\right)_{2} + \left(\begin{array}{c} 
\{A_{y}B_{z}\}\\
\{A_{z}B_{x}\}\\
\{A_{x}B_{y}\} \end{array}\right)_{3} 
+ \left(\begin{array}{c} 
[A_{y}B_{z}]\\
\left[A_{z}B_{x}\right]\\
\left[A_{x}B_{y}\right] \end{array}\right)_{3^{\prime}}.     
\end{gather}
\begin{equation}
\begin{gathered}
A \cdot B = A_{x}B_{x}+ A_{y}B_{y} +  A_{z}B_{z}\\ 
\{A_{i}B_{j}\} = A_{i}B_{j} + B_{j}A_{i}\\ 
\left[A_{i}B_{j}\right] = A_{i}B_{j} - A_{j}B_{j}\\   
A \cdot \Sigma \cdot B = A_{x}B_{x} + \omega A_{y}B_{y} + \omega^2 A_{z}B_{z}\\  
A \cdot \Sigma^{*} \cdot B = A_{x}B_{x} + \omega^2 A_{y}B_{y} + \omega A_{z}B_{z}.  
\end{gathered}
\end{equation}
Later on for simplicity, we can replace $3 \rightarrow 3_{1}$, $3^{\prime} \rightarrow 3_{2}$, $1 \rightarrow 1_{1}$, 
$1^{\prime} \rightarrow 1_{2}$.
\begin{gather*}
2 \otimes 3_{1}= 3_{1} \oplus 3_{2},\\
3_{1} \otimes 3_{1}= 1_{1} \oplus 2 \oplus 3_{1} \oplus 3_{2},\\
3_{2} \otimes 3_{2}= 1_{1} \oplus 2 \oplus 3_{1} \oplus 3_{2}.
\end{gather*}
The Clebsch-Gordon coefficients for $2 \otimes 3_{1}$, used in our analysis is as follows    
\begin{gather*}
\left(\begin{array}{c}
a_{1}\\
a_{2}\end{array}\right)_{2}\otimes \left(\begin{array}{c}
b_{1}\\
b_{2}\\
b_{3} \end{array}\right)_{3_{1}} = \left(\begin{array}{c}
a_{2}b_{1}\\
-\frac{1}{2}(\sqrt{3}a_{1}b_{2}+a_{2}b_{2})\\
\frac{1}{2}(\sqrt{3}a_{1}b_{3}-a_{2}b_{3}) \end{array}\right)_{3_{1}} \oplus \left(\begin{array}{c}
a_{1}b_{1}\\
\frac{1}{2}(\sqrt{3}a_{2}b_{2}-a_{1}b_{2})\\
-\frac{1}{2}(\sqrt{3}a_{2}b_{3}+a_{1}b_{3}) \end{array}\right)_{3_{2}}.
\end{gather*}
Similarly,the Clebsch-Gordon coefficients for $3_{1} \otimes 3_{1}$, used in our analysis can be expressed as,    
\begin{gather*}
\left(\begin{array}{c}
a_{1}\\
a_{2}\\
a_{3} \end{array}\right)_{3_{1}}\otimes \left(\begin{array}{c}
b_{1}\\
b_{2}\\
b_{3} \end{array}\right)_{3_{1}} = (a_{1}b_{1}+a_{2}b_{2}+a_{3}b_{3})_{1_{1}} \oplus \left(\begin{array}{c}
1/\sqrt{2}(a_{2}b_{2} - a_{3}b_{3})\\
1/\sqrt{6}(-2a_{1}b_{1}+ a_{2}b_{2} + a_{3}b_{3})
\end{array}\right)_{2} \oplus  \\
\left(\begin{array}{c}
a_{2}b_{3}+a_{3}b_{2}\\
a_{1}b_{3}+a_{3}b_{1}\\
a_{1}b_{2}+a_{2}b_{1} \end{array}\right)_{3_{1}} \oplus \left(\begin{array}{c}
a_{3}b_{2}-a_{2}b_{3}\\
a_{1}b_{3}-a_{3}b_{1}\\
a_{2}b_{1}-a_{1}b_{2} \end{array}\right)_{3_{2}}.
\end{gather*}
The Clebsch-Gordon coefficients for $3_{2} \otimes 3_{2}$, used in our analysis is given as,    
\begin{gather*}
\left(\begin{array}{c}
a_{1}\\
a_{2}\\
a_{3} \end{array}\right)_{3_{2}}\otimes \left(\begin{array}{c}
b_{1}\\
b_{2}\\
b_{3} \end{array}\right)_{3_{2}} = (a_{1}b_{1}+a_{2}b_{2}+a_{3}b_{3})_{1_{1}} \oplus \left(\begin{array}{c}
1/\sqrt{2}(a_{2}b_{2} - a_{3}b_{3})\\
1/\sqrt{6}(-2a_{1}b_{1}+ a_{2}b_{2} + a_{3}b_{3})
\end{array}\right)_{2} \oplus  \\
\left(\begin{array}{c}
a_{2}b_{3}+a_{3}b_{2}\\
a_{1}b_{3}+a_{3}b_{1}\\
a_{1}b_{2}+a_{2}b_{1} \end{array}\right)_{3_{1}} \oplus \left(\begin{array}{c}
a_{3}b_{2}-a_{2}b_{3}\\
a_{1}b_{3}-a_{3}b_{1}\\
a_{2}b_{1}-a_{1}b_{2} \end{array}\right)_{3_{2}}.
\end{gather*}
\section{Derivation of Boltzmann Equation in FRW Universe:Expression for relic density}
\label{appen2}
Boltzmann equation is integral partial differential equation for the phase space distribution of all the species present in the Universe. The evolution of phase space distribution function of the particles is governed by Boltzmann equation and solution of this equation gives the relic density of the species. Solution of Boltzmann equation is important to  find the relic abundance of the dark matter candidate. The number density is governed by\cite{steigman2012precise}
\begin{equation}\label{1}
\frac{dn_{a}}{dt}+ 3Hn_{a} = C_{a}
\end{equation}
\\where $C_{a}$ is the collision operator representing all interactions present among the particles and H is the Hubble parameter.
We consider $2\longrightarrow2$ process in which $C_{a}$ can be written as\cite{d2010semi}
\begin{equation}\label{eq:a2}
{C}_{ab\longrightarrow cd} = - \int(2\pi)^{4}\delta^{4}(p_{a}+p_{b}-p_{c}-p_{d})d\pi_{a}d\pi_{b}d\pi_{c}d\pi_{d} [|M_{ab\longrightarrow cd}|^{2}f_{a}f_{b} -|M_{cd\longrightarrow ab}|^{2}f_{c}f_{d}
\end{equation}
where $d\pi_{i}$ represents the phase space and the expression for $d\pi_{i}$ obtained as follows,
\begin{equation}\label{eq:a3}
d\pi_{i} = g_{i}\frac{d^3\pi}{2E\times(2\pi)^{3}}
\end{equation}
and,
\begin{equation}\label{eq:a4}
f_{i} = \frac{n_{i}(t)}{n_{i}^{eq}(t)}f_{i}^{{eq}}(E,t)
\end{equation}

Here,we consider CP or T invariance so that $$|M_{ab\longrightarrow cd}|^{2} = |M_{cd\longrightarrow ab}|^{2} =|M|^{2}$$ and equation \eqref{eq:a2} can be read as,
\begin{equation}\label{eq:a10}
{C}_{ab\longrightarrow cd} = - \int(2\pi)^{4}\delta^{4}(p_{a}+p_{b}-p_{c}-p_{d})d\pi_{a}d\pi_{b}d\pi_{c}d\pi_{d} [|M|^{2}(f_{a}f_{b} -f_{c}f_{d})]
\end{equation}
$(f_{a}f_{b} -f_{c}f_{d})$ is the collision term in the Boltzmann equation.
\begin{equation}\label{eq:a11}
f_{i} = exp[-\frac{E_{i}-\mu}{T}]
\end{equation}
We assume chemical potential to be zero $(\mu=0)$  which gives $ (f_{a}f_{b} -f_{c}f_{d})$ = $[f_{a}f_{b} - f_{a}^{eq}f_{b}^{eq}]$ as energy part of the delta function enforces$E_{a}+E_{b} = E_{c}-E_{d}$, using equation\eqref{eq:a4}we get,
\begin{equation}\label{eq:a12}
{C}_{ab\longrightarrow cd} = -\int  \frac{1}{{n_{a}}^{eq}{n_{b}}^{eq}}(2\pi)^{4}\delta^{4}(p_{a}+p_{b}-p_{c}-p_{d})d\pi_{a}d\pi_{b}d\pi_{c}d\pi_{d}|M|^{2}exp[-\frac{E_{a}}{T}]exp[-\frac{E_{b}}{T}][n_{a}n_{b} -n_{a}^{eq}n_{b}^{eq}]
\end{equation}
Again considering $n_{a} = n_{b}$  we can write,
\begin{equation}\label{eq:a5}
C_{ab\longrightarrow cd} = -\int \frac{1}{({n_{a}}^{eq})^{2}}(2\pi)^{4}\delta^{4}(p_{a}+p_{b}-p_{c}-p_{d})d\pi_{a}d\pi_{b}d\pi_{c}d\pi_{d} |M|^{2}exp[-\frac{E_{a}}{T}]exp[-\frac{E_{b}}{T}][n_{a}^{2}-(n_{a}^{eq})^2]
\end{equation}
\begin{equation}\label{eq:final}
{C}_{ab\longrightarrow cd} = {<\sigma|v>}_{rel}[n_{a}^{2}-(n_{a}^{eq})^2]
\end{equation}
where,the thermally averaged annihilation cross-section is given as,$${<\sigma|v>}= -\int \frac{1}{({n_{a}}^{eq})^{2}}(2\pi)^{4}\delta^{4}(p_{a}+p_{b}-p_{c}-p_{d})d\pi_{a}d\pi_{b}d\pi_{c}d\pi_{d} |M|^{2}exp[-\frac{E_{a}}{T}]exp[-\frac{E_{b}}{T}]$$\\Thus the relic abundance of a any particle $\chi$ is governed by the Boltzmann equation as follows
\begin{equation}\label{eq:13}
\frac{dn_{\chi}}{dt}+ 3Hn_{\chi} = {<\sigma|v>}_{rel}[n_{\chi}^{2}-(n_{\chi}^{eq})^2]
\end{equation}
To solve it, we replace the variable number density n by $Y = \frac{n}{S}$, S being the entropy of the Universe and $$t= 0.301 g^{-2}\frac{m_{pl}}{m^{2}}x^{2}$$.  Now we have
\begin{equation}\label{eq:final2}
\frac{dY}{dx} = -\frac{x<\sigma|v>S}{H(m)}(Y^{2}-{Y_{eq}}^{2})
\end{equation}
The relic abundance of a given elementary particle is a measure of the present quantity of that particle remaining from the Big Bang. It can be expressed as\cite{kolb2018early},
\begin{equation}
\Omega_{\chi}h^{2} =\frac{\rho_{0}}{\rho_{crit}}= \frac{S_{0}Y_{\infty} m}{\rho_{crit}}
\end{equation}
where $S_{0}$ presents entropy and$Y_{\infty} = Y(x=\infty)$ is the present value. $\rho_{crit}$ can be obtained from $8\Pi G \rho_{crit}= 3{H_{0}}^{2}$, ${H_{0}}$ is the Hubble constant. Thus ,solving the equation \eqref{eq:final2} for $Y(x =\infty)$, one can get the expression for relic abundance of a species.
In case of sterile neutrinos, the relic abundance can be expressed as\cite{asaka2007lightest}:
\begin{equation}
\Omega_{\alpha s} = \frac{m_{s}Y_{\alpha s}}{3.65\times10^{-9}h^{2} GeV}
\end{equation}
Here, $m_{s}$ represents the mass of the sterile neutrino and $Y_{\alpha s}$ is the neutrino Yukawa couplings. Using the parametrisation $|\epsilon_{\alpha s}| \approx \theta_{\alpha s}$\cite{abada2014dark},we get,
\begin{align}
\Omega_{\alpha s} h^{2} &= 0.273\times10^{9}\frac{m_{s}Y_{\alpha s}{|\epsilon_{\alpha s}|}^{2}}{ \theta_{\alpha s}^{2}}GeV^{-1}\nonumber \\
&=0.11\times2.49\times10^{-5}\frac{m_{s}Y_{\alpha s}}{{\theta_{\alpha s}^{2}}}(\frac{|\epsilon_{\alpha s}|}{0.1 eV})^{2}\nonumber \\
&= 1.1\times10^{7}C_{\alpha}(m_{s})|\epsilon_{\alpha s}|(\frac{m_{s}}{keV})^{2}
\end{align}
where,
\begin{equation}
C_{\alpha}(m_{s}) = 2.49\times10^{-5}\frac{Y_{\alpha s}}{\theta_{\alpha s}^{2}}\frac{keV}{m_{s}}
\end{equation}
\section*{Acknowledgements}
NG acknowledges Department of Science and Technology (DST),India(grant DST/INSPIRE Fellowship/2016/IF160994) for the financial assistantship. The work of MKD is supported by the Department of Science and Technology, Government of India under the project no. $EMR/2017/001436$.
\bibliographystyle{paper}
\bibliography{SterileDM}
\end{document}